%% file: pair_stat_final.tex
\documentclass[prc,aps,groupedaddress,superscriptaddress,twocolumn,nofootinbib,showpacs,showkeys,floatfix,a4paper,10pt]{revtex4-1}

\usepackage[colorlinks,citecolor=blue]{hyperref}
\usepackage{graphicx,colordvi,rotating}
\usepackage{multirow,color}
\usepackage{amsmath,amssymb}
\usepackage[mathscr]{euscript}
\usepackage{cancel}
\usepackage{braket}
\usepackage{cases}
\usepackage{color}
\usepackage{mathtools}
\usepackage{natbib}
\usepackage{ulem}
\usepackage{graphicx}% Include figure files
\usepackage{dcolumn}% Align table columns on decimal point
\usepackage{bm}% bold math
\usepackage{natbib}
\usepackage{enumerate}
\usepackage[dvipsnames]{xcolor}
\usepackage{amsmath}
\usepackage{wrapfig}
\usepackage[utf8]{inputenc}
\usepackage{epstopdf}

\usepackage{t1enc}
\usepackage[utf8]{inputenc}

\begin{document}

%\preprint{APS/123-QED}

\title{Pairing in Fission: Mean-Field and Collective Inertias Study}

\author{A.~Zdeb}
\email{azdeb@kft.umcs.lublin.pl}
\affiliation{Institute of Physics, Maria Curie-Skłodowska Univeristy, Lublin, Poland}

\author{M.~Warda}
\email{michal.warda2@mail.umcs.pl}
\affiliation{Institute of Physics, Maria Curie-Skłodowska Univeristy, Lublin, Poland}

\author{L.~M.~Robledo}
\email{luis.robledo@uam.es}
\affiliation{Departamento de Física Teórica and CIAFF, Universidad Autónoma de Madrid, Madrid, Spain}
\affiliation{Center for Computational Simulation, Universidad Politécnica de Madrid, Campus de Montegancedo, Bohadilla del Monte, E-28660 Madrid, Spain}

\author{S.~A.~Giuliani}
\email{samuel.giuliani@uam.es}
\affiliation{Departamento de Física Teórica and CIAFF, Universidad Autónoma de Madrid, Madrid, Spain}
\date{\today}% It is always \today, today,
%\author{A. Baran}
%\affiliation{Institute of Physics, Maria-Curie-Skłodowska University, Lublin, Poland}

\begin{abstract}
Pairing plays a crucial role in the microscopic description of nuclear 
fission. Microscopic methods provide access to three quantities related 
to pairing, namely, the pairing gap ($\Delta$), the particle number 
fluctuations ($ \Delta \hat{N}^2 $), and the quenching factor (QF) of the pairing strength. The 
aim of this work is to analyse the impact of each of these quantities 
on the mean-field description of the fission process, including collective
inertias. 
\end{abstract}


%Use showkeys class option if keyword
\maketitle

%%%%%%%%%%%%%%%%%%%%%%%%%%%%%%%%%%%%%%%%%%%%%%%%%
%%%%%%%%%%%%%%%%%%%%%%%%%%%%%%%%%%%%%%%%%%%%%%%%%
%%%%%%%%%%%%%%%%%%%%%%%%%%%%%%%%%%%%%%%%%%%%%%%%%
%%%%%%%%%%%%%%%%%%%%%%%%%%%%%%%%%%%%%%%%%%%%%%%%%
%%%%%%%%%%%%%%%%%%%%%%%%%%%%%%%%%%%%%%%%%%%%%%%%%
%%%%%%%%%%%%%%%%%%%%%%%%%%%%%%%%%%%%%%%%%%%%%%%%%

%%%%%%%%%%%%%%%%%%%%%%%%%%%%%%%%%%%%%%%%%%%%%%%%%
\section{Introduction\label{INTRODUCTION}}
%%%%%%%%%%%%%%%%%%%%%%%%%%%%%%%%%%%%%%%%%%%%%%%%%

In spite of having been discovered more than 80 years ago ~\cite{HS39}, 
the nuclear fission process still remains one of the most fascinating 
and not fully understood nuclear 
phenomenon~\cite{Schmidt2018,Bender2020}. Shortly after the first 
observation of fission in the laboratory, the process was explained 
using the liquid drop model assumption of the nucleus \cite{MF39} and 
this interpretation is still valid nowadays. The subsequent discovery 
of spontaneous fission~\cite{FP40}, indicated that the 
fission process itself is much more complex than initially thought. 
These initial studies opened up the door to a myriad of fascinating 
experimental and theoretical studies that have been carried out, and 
published since then.

It is widely recognized that the non-negligible role of pairing correlations 
in fission dynamics strongly affects the calculation of 
half-lives.  Two main factors play a crucial role in this connection. 
First, the modification of the fission barrier height and nuclear 
deformation at the barrier induced by pairing. Second,  the reduction 
of the collective inertia along the path connecting the ground state 
with the scission point. These two aspects have been extensively 
studied in the literature, employing a large variety of nuclear models 
and interactions.

The first study on the impact of pairing correlations on fission was 
conducted in the early 1970s. Brack et al. noticed that mass parameter 
in fission process follows almost exactly a $1/\Delta^2$ law (here 
$\Delta$ is the pairing gap) due to the impact on level density near the 
Fermi energy induced by pairing \cite{Brack1972}. They realized that 
fission half-lives may be significantly lowered when the fission path does 
not follow the lowest energy path in the potential energy, and it does 
not necessarily pass through the extremal points of the deformation 
energy. Soon after, Ledergerber and Pauli~\cite{LEDERGERBER19731} 
determined fission paths using the liquid drop model with Strutinski 
shell correction and (surface) pairing treated in the 
Bardeen-Cooper-Schrieffer (BCS) approach. The authors noticed the 
strong dependence of the inertia parameter with respect to the value of 
pairing gap $\Delta$ in the BCS model. In consequence,  the 
consideration of pairing leads to the reduction of the tunneling 
probability through the fission barrier without modifying the potential 
energy.  

Moreto and Babinet ~\cite{babinet} applied a simple, realistic model to 
show that an increase in the pairing gap parameter leads to an increase 
in the penetrability of the fission barrier. The main message from this 
study is that there is a strong interplay between the potential energy, 
which increases as a square of the pairing gap, and the collective 
inertia, which decreases as an inverse of the pairing gap squared. As a 
consequence, the least-action path leads through regions with a larger 
pairing gap than the least-energy path.

The study of the influence of pairing on fission observables was 
extensively explored in a series of papers coming from the nuclear 
physics group at Maria Curie-Sklodowska University in 
Lublin~\cite{STASZCZAK1985227, LOJEWSKI1980134,  STASZCZAK1989589, 
LOJEWSKI1999}. The fission half-lives were computed using various 
macroscopic-microscopic approaches, including pairing treated in the BCS 
framework. Proton ($\Delta_p$) and neutron ($\Delta_n$) pairing gap 
parameters were treated on equal footing as the standard shape 
deformation parameters.  Two types of fission paths were considered: 
{\it static} (also known as {\it least-energy}) path as a result of the 
minimization of the energy on the way from ground state to 
scission and {\it dynamic} (also named {\it least-action}) path where 
the action integral along the path is minimized. The second one is more 
demanding computationally as it requires simultaneous analysis of both 
energy and collective inertia at each point of the space of 
(deformation) parameters. In the mentioned papers, the dynamical 
fission path was determined by minimizing the action integral in the 
multidimensional space that included pairing degrees of 
freedom~\cite{STASZCZAK1985227,  LOJEWSKI1999}. This is because of the 
strong dependence of the inertia with the pairing gap discussed above. 
As an example, it was shown that in $^{252}$Fm, the shortest half-lives 
are obtained for the fission path with  $\Delta$ values larger than 
those in the static path minimizing the energy.  In the dynamic path 
the height of the fission barrier increases by 1.4 MeV with respect to 
the result for the static path but the mass parameter along the dynamic 
path gets reduced by a factor of two, leading to a reduction of 
half-lives of three orders of magnitude. A similar effect was observed 
for other members of the fermium isotopic chain. 

In the previous papers, only the monopole component of pairing was 
considered. The impact of quadrupole pairing on the spontaneous fission 
half-lives of Fm isotopes was studied in~\cite{LOJEWSKI1980134} in the 
static approach. Quadrupole pairing decreases the height of the second 
barrier by 1 MeV, whereas the mass parameter tends to increase, 
especially on the first fission barrier. Overall, including 
quadrupole pairing leads to a reduction in the half-lives of up to two 
orders of magnitude in some of the Fm isotopes considered.

In Ref.~\cite{STASZCZAK1989589}, the generator coordinate method, 
together with the Gaussian overlap approximation (GCM+GOA), was applied 
to investigate the influence of pairing vibrations on the spontaneous 
fission half-lives calculations of even-even transplutonium isotopes. 
The authors found that the action integrals calculated along the static 
fission paths are around 13\% larger 
than the integrals obtained from dynamical trajectories. This is a 
consequence of the decrease of mass parameters, which compensates the 
increase of fission barriers. Considering dynamical fission paths with 
pairing as a collective variable reduces the calculated half-lives by 
several orders of magnitude. The authors of Ref. \cite{KPIV09} noticed that deformation dependent pairing energy
decreases barrier heights in light nuclei improving agreement with experimental data.

Previous findings are also observed in studies using self-consistent 
mean field methods. In the framework of the Relativistic Hartree-Bogoliubov (RHB) and
Relativistic Mean Field (RMF) plus BCS methods it was shown in Ref.~\cite{KARATZIKOS201072} 
the strong dependence of pairing gaps with deformation (invalidating the 
assumption of using a constant gap) as well as the impact on the barrier
heights depending on the use of zero or finite range pairing forces.

In Refs.~\cite{Giuliani2014,PhysRevC.90.061304, PhysRevC.93.044315, 
PhysRevC.107.044307}, the impact of pairing on fission was studied 
using the particle-number dispersion $\Delta \hat{N}^2$ as collective 
degree of freedom.  These studies, which involved Skyrme, Gogny, and 
relativistic energy density functionals, showed that the inclusion of 
$\Delta \hat{N}^2$ on the same footing as deformation parameters 
strongly modifies the 
least-action paths, producing an overall reduction of spontaneous 
fission lifetimes as compared to static calculations and dynamical 
calculations including only mass multipole collective degrees of 
freedom \cite{PhysRevC.88.064314}.  Additionally, it has been 
demonstrated that in some nuclei pairing as a dynamic degree of freedom 
restores the axial symmetry at the first fission 
barrier~\cite{PhysRevC.90.061304, PhysRevC.93.044315}.  Time-dependent 
GCM+GOA calculation employing $\Delta \hat{N}^2$ as collective degree 
of freedom have shown that pairing correlations play an important role 
in shaping the fission fragment yields in induced 
fission~\cite{PhysRevC.104.044612}. Also, $\Delta \hat{N}^2$  has been 
employed in beyond-mean field studies exploring the role of dynamic 
pairing correlations in fission~\cite{PhysRevC.107.044307}.

The strength of pairing interactions, whether monopole, zero range, finite range, etc
is usually fitted to reproduce odd-even staggering data on binding energies.
The physics of staggering is far away from the physics of fission, and therefore
one may wonder whether pairing strengths could be fine-tuned specifically 
to reproduce fission observables like spontaneous fission lifetimes. To
explore this possibility, it is quite common to analyze the behavior of
fission observables as a function of some parameter modulating pairing
strength.  In a recent example ~\cite{Kouno22,  10.1093taiki}, where the pairing
strength in an RMF+BCS calculation was increased, a reduction of
up to 2.5 MeV in the barrier height of several Pu isotopes was observed. The
reduction is the consequence of a larger binding energy gain for the ground
state than for the top of the barrier. 
This result is in contradiction with the research presented above when 
dynamic fission paths are characterized by stronger pairing and higher 
barriers at the same time. A similar result was found in
Refs.~\cite{PhysRevC.104.044329,PhysRevC.107.034307}, using a  
macroscopic-microscopic model with BCS monopole pairing.
These results were confirmed in energy density functional studies
employing the Skyrme forces in the fission of $^{240}$Pu~\cite{Chen_2022}. There,
a decrease of the fission barriers up to 2.3~MeV in both symmetric and asymmetric fission 
channels is observed when the 
pairing strength is increased by 10\%. The deformation of the nucleus at the scission 
line can also be slightly modified by modifications in the pairing interaction strength.
A similar result was also discussed in Ref. ~\cite{PhysRevC.108.034306}. 
In microscopic Hartree-Fock-Bogoliubov (HFB) calculations with a finite range pairing force, 
the impact of varying the pairing strength was analyzed in Refs.~\cite{Giuliani2013, Rodriguez-Guzman2014, Rodriguez-Guzman2014a,guzman17}.
In those studies, increasing the pairing strength by 5\%-10\% can reduce
the spontaneous fission lifetimes by several orders of magnitudes, particularly
for those nuclei with broad and high fission barriers, mostly because of the
strong damping of the collective inertias. The effect is particularly important to
explain the odd-even staggering in spontaneous fission lifetimes \cite{guzman17}. Moreover, when pairing is stronger, the mass parameters show 
smoother behavior without sharp peaks along the fission path. 

All the above concerns mean field pairing correlations with their inherent 
spontaneous symmetry-breaking mechanism, and therefore, they do not take 
into account the impact of restoring particle number symmetry. 
Typically, symmetry restoration effects in the variation after 
projection (VAP) scheme also tend to increase pairing correlations 
\cite{PhysRevC.99.064301,PhysRevC.106.024335} but the effect is partly compensated by the
reduction of pairing correlations associated with Coulomb antipairing, a
component of the Coulomb interaction that is usually not considered.

As the short literature overview presented here shows, the role of 
pairing in fission has been extensively studied in the last fifty 
years. The non-negligible impact of pairing in fission barriers and 
half-lives is currently well known. However, as various theoretical 
methods were used to characterize the importance of pairing 
correlations, results obtained from different studies cannot always be 
directly compared. Moreover, some contradictory conclusions were drawn 
from these studies,  namely differences in the modification of the size 
of the fission barrier with pairing.  These ambiguities indicate the 
need to systematize the knowledge regarding the impact of pairing in 
static fission calculations. In this work, we closely examine the three 
self-consistent model parameters corresponding to pairing and 
thoroughly investigate their impact on the description of the fission 
process, employing a consistent theoretical framework based on the HFB 
method.  The paper is organized as follows: Section~\ref{THEORY} 
briefly presents the theoretical model employed in this work. 
Section~\ref{RESULTS} presents the results of our studies. We first 
address the impact of the pairing gap, the particle number 
fluctuations, and the quenching factor parameters on fission barriers. 
We then discuss the results of fission half-lives for various pairing 
strengths and different aspects of collective inertia parameters along 
fission paths. In Section~\ref{CONCLUSIONS}, we summarize our main 
conclusions. In Appendix~\ref{APPENDIX}, we discuss the problem of the 
relative sign of the occupancies in the Bogoliubov transformation and 
its impact in the calculations present here.

%%%%%%%%%%%%%%%%%%%%%%%%%%%%%%%%%%%%%%%%%%%%%%%%%
\section{Theoretical framework \label{THEORY}}
%%%%%%%%%%%%%%%%%%%%%%%%%%%%%%%%%%%%%%%%%%%%%%%%%

The starting point for our work is the computation of potential energy 
surface (PES) modeling fission. We use the microscopic, 
self-consistent HFB model with the finite-range Gogny-type nuclear 
interaction. We applied the widely used and well-established D1S 
parametrization~\cite{D1S1, D1S2, D1S3, D1S4, D1S5}. The computations 
are done in a one-center axially symmetric, deformed harmonic oscillator 
basis with 16 shells (including as many as 23 one-dimensional harmonic 
oscillator shells along the symmetry axis $n_{z\,\textup{max}}=22$). 
The spontaneous fission half-lives are computed using the 
Wentzel-Kramers-Brillouin (WKB) approximation, assuming the axially 
symmetric quadrupole moment $Q_{20}$ as the leading coordinate. The 
estimation of the probability of tunneling through the one-dimensional 
fission barrier requires the computation of the action $S$ along the 
least-energy path~\cite{Brack1972}:
\begin{equation}
	S=\int_{\textup{in}}^{\textup{out}}\sqrt{2B(Q_{20})(V(Q_{20})-E_0)}dQ_{20} \,,
\label{s}
\end{equation}
being $\textup{in}$ and $\textup{out}$ the classical turning points 
defined by $V(Q_{20})=E_0$.  The potential energy appearing in the expression of 
the action $V(Q_{20})$ is the sum of the HFB energy and the rotational 
and zero point energy correction~\cite{Egido2004, RS80}. The $E_0$ parameter represents the 
quantal ground state energy obtained by solving the collective 
Schr\"odinger equation with the quadrupole moment as the degree of freedom. 
Although this parameter can be easily estimated by assuming a parabolic 
behavior of the PES around the ground state minimum and considering the 
collective inertia at that point, we employ a more pragmatic approach 
and take it equal to 0.5 MeV, which is a reasonable value.
$B(Q_{20})$ is the collective inertia calculated within the 
perturbative {\it cranking} approximation~\cite{PhysRevC.84.054321}:
\begin{equation}
	B(Q_{20})=\frac{M_{-3}(Q_{20})}{M_{-1}^2(Q_{20})} \,,
\label{b}
\end{equation}
where  $M_{-n}$ are moments of order $n$ of the 
collective coordinate $Q_{20}$ (see~\cite{PhysRevC.84.054321} for the 
definition). The formula for the inertia is based on the assumption that
the collective operator is a one-body one, which is not true for
the fluctuation on the number of particles operator $\Delta\hat{N}^{2}$ used below.
To overcome this problem, the traditional approach is
to use in the expression of the moments the one-body part  of
$\Delta\hat{N}^{2}$ in the quasiparticle basis. Although this seems a 
reasonable assumption, it is desirable to test it against the results
obtained from a truly one-body operator, like the pairing gap one discussed
below.
Within the WKB approximation, the spontaneous fission half-life (in seconds) is
given by~\cite{Schunck2016}:
\begin{equation}
t_{sf}=2.86\times 10^{-21}(1 + \exp(2S)). 
\label{t}
\end{equation}

As mentioned in the Introduction~\ref{INTRODUCTION}, the goal of the present work is to 
discuss the impact of pairing correlations on the quantities
entering the action integral~\eqref{b}.
To do so, our analysis covers the three main methods that have been traditionally
employed to characterize pairing correlations strengths in fission:
HFB calculations with constraints on the pairing gap  parameter ($\Delta$),
on the particle number fluctuation operator ($\Delta \hat{N}^2$), and the
renormalization of the pairing interaction strength. 

The pairing gap operator can be defined as $\Delta = 
\frac{G}{2}(\hat{P}+\hat{P}^+)$, being  $\hat{P}^+$ the Cooper pair 
creation operator, and $G=10/A$ (MeV), a parameter related to the 
typical strength of the monopole pairing interaction. By employing 
$\Delta$ as a constraining operator, it is thus possible to control the 
amount of pairing correlations in HFB wave functions $|\Phi\rangle$. 
The main advantage of employing $\Delta$ as a collective degree of 
freedom is its one-body nature, which allows a consistent estimation of 
collective inertias through Eq.~\eqref{b}. However, in order to employ 
$\Delta$ as a collective degree of freedom, specific numerical 
strategies must be followed in order to properly evaluate the gradient 
of the energy and other expectation values, as detailed in the 
Appendix~\ref{APPENDIX}.

As one of the consequences of the existence of pairing correlations in 
a HFB calculation is the spontaneous symmetry breaking of the particle 
number symmetry, it is possible to employ one of the associated order 
parameters as a measure of pairing correlations. The most 
straightforward order parameter is the particle number fluctuation 
$\Delta \hat{N}^2 = (\hat{N} -  \langle \hat{N} \rangle)^2$, being 
$\hat{N}$ the particles number operator. This quantity is zero in the 
absence of pairing correlations and becomes relatively large (of the 
order of a few units) for strongly correlated nuclear systems. As 
mentioned before, a drawback related to the fluctuation in particle 
number is its two-body-operator character. However, owing to the 
gradient method used in the solutions of the HFB equation, two-body 
constraint operators can be handled on the same footing as one-body 
ones.

As a last quantity characterizing pairing correlations, we consider the quenching factor (QF),
a parameter that renormalizes the strength of the pairing 
interaction of the Gogny force. The QF 
is introduced by multiplying the HFB pairing field $\Delta^{(\tau)}$ 
($\tau=p,n$) by a constant factor in each iteration of the iterative 
solution of the HFB equation. This QF parameter is equal to 1.0 for not-modified 
HFB fields and takes typical values around 1.1 for 
enhanced pairing. It can be pictured as an analog of the pairing 
strength $G$ of other simple models, like BCS\@. In calculations employing the
QF, pairing correlations can be characterized by the particle-particle energy 
$E_{pp}$ for protons and neutrons, which in HFB formalism is given by:
\begin{equation}
E_{pair}^{(\tau)}=\frac{1}{2}\rm{Tr}(\Delta^{(\tau)}\kappa^{(\tau)}),
\label{epairing}
\end{equation}
where $\Delta^{(\tau)}$ is the pairing field and $\kappa^{(\tau)}$ 
stands for the pairing tensor obtained from the solution of the HFB 
equations. Although this quantity is not the pairing energy of the 
system (i.e., the energy difference between the paired solution and the 
corresponding Hartree-Fock energy), one usually refers to it as 
``pairing energy", and we will follow in subsequent sections this 
custom.

%%%%%%%%%%%%%%%%%%%%%%%%%%%%%%%%%%%%%%%%%%%%%%%%%
\section{RESULTS\label{RESULTS}}
%%%%%%%%%%%%%%%%%%%%%%%%%%%%%%%%%%%%%%%%%%%%%%%%%s
%

%\input{../../262Rf/n2/d1s_qf1/deltaodn2}
\begin{figure}[h!]
\hskip-0.5cm
\includegraphics[scale=0.136,angle=0]{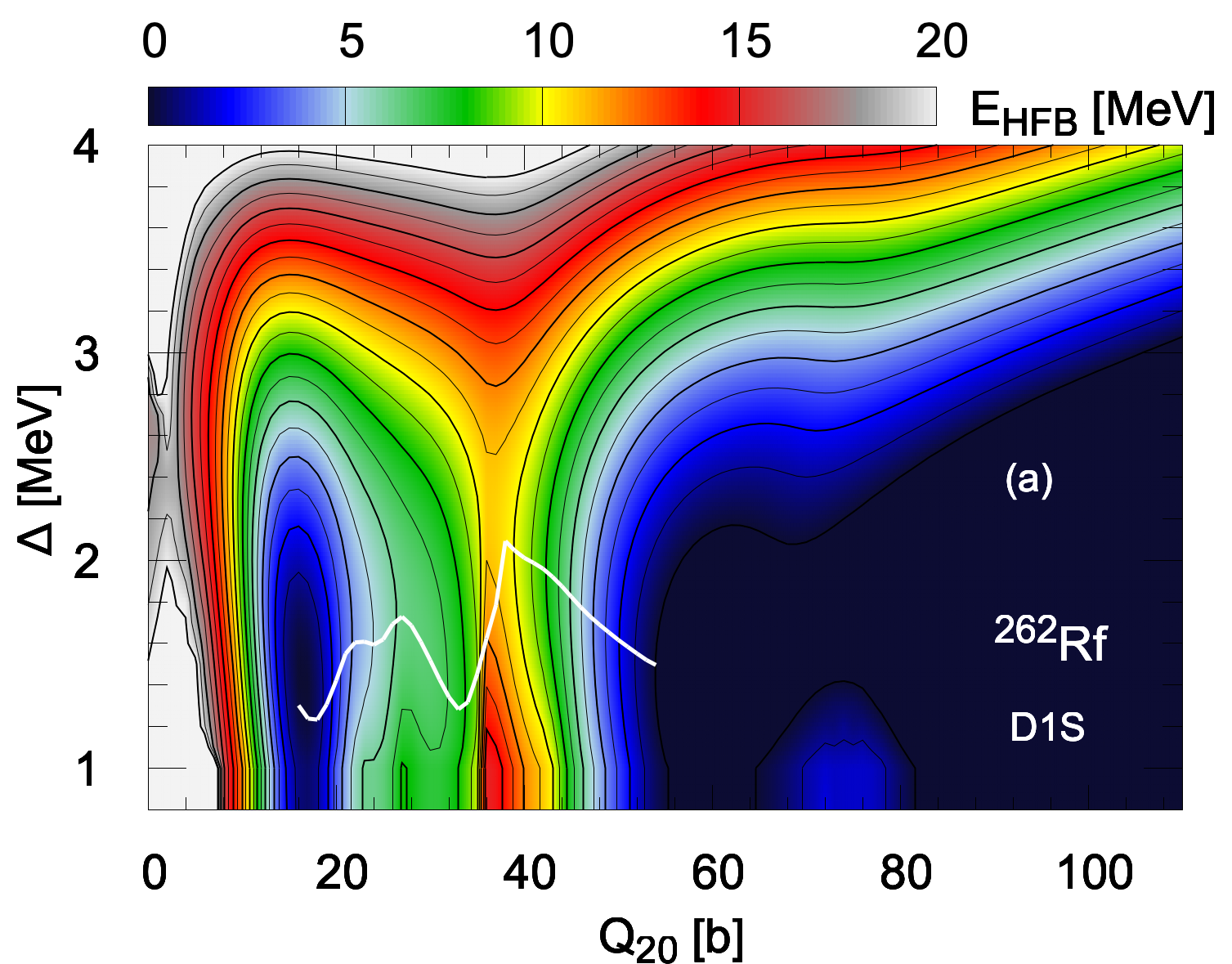}\\
\vskip-0.05cm
\hskip-0.5cm
\includegraphics[scale=0.138,angle=0]{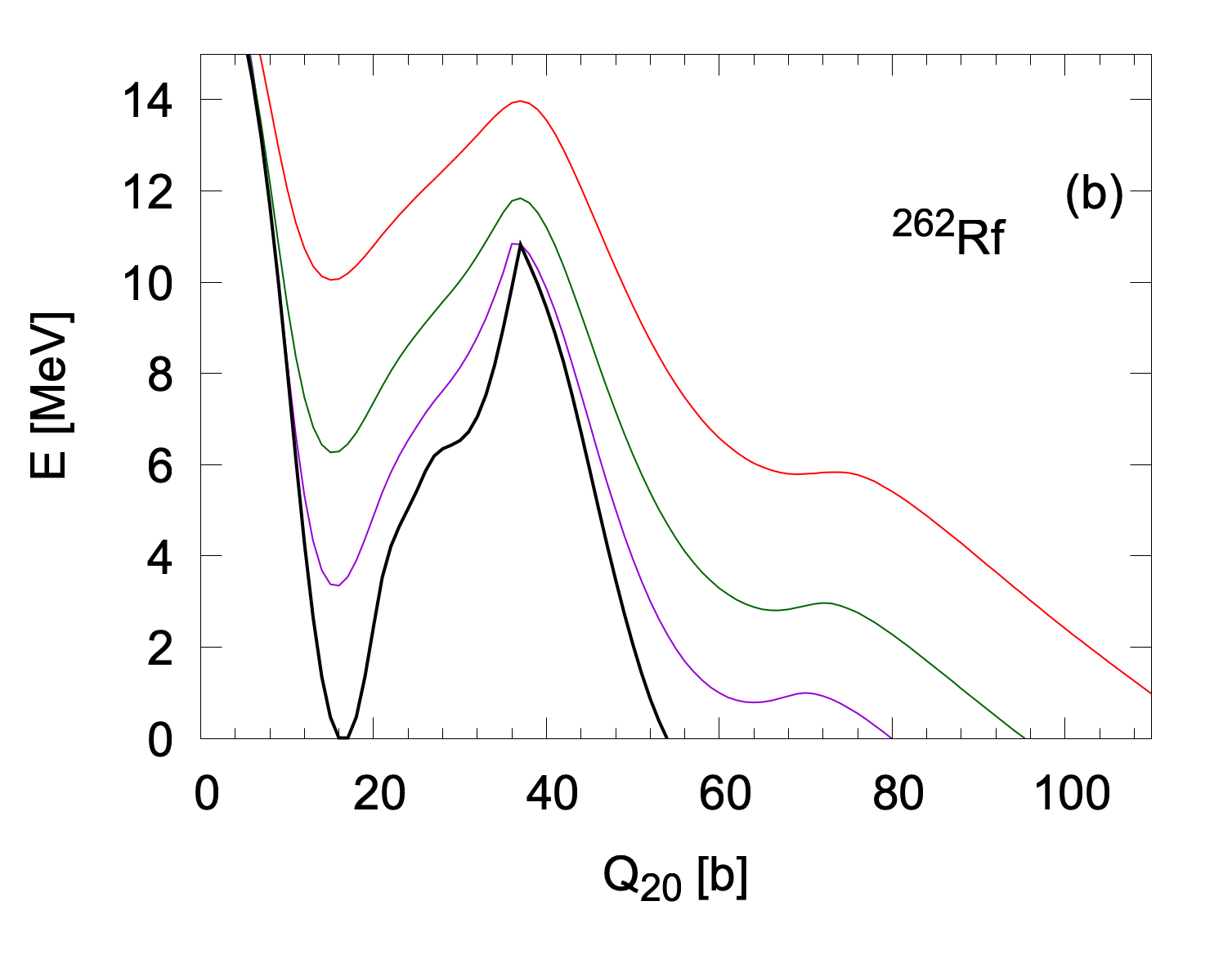}\\
\vskip-1.2cm
\hskip-0.5cm
\includegraphics[scale=0.138,angle=0]{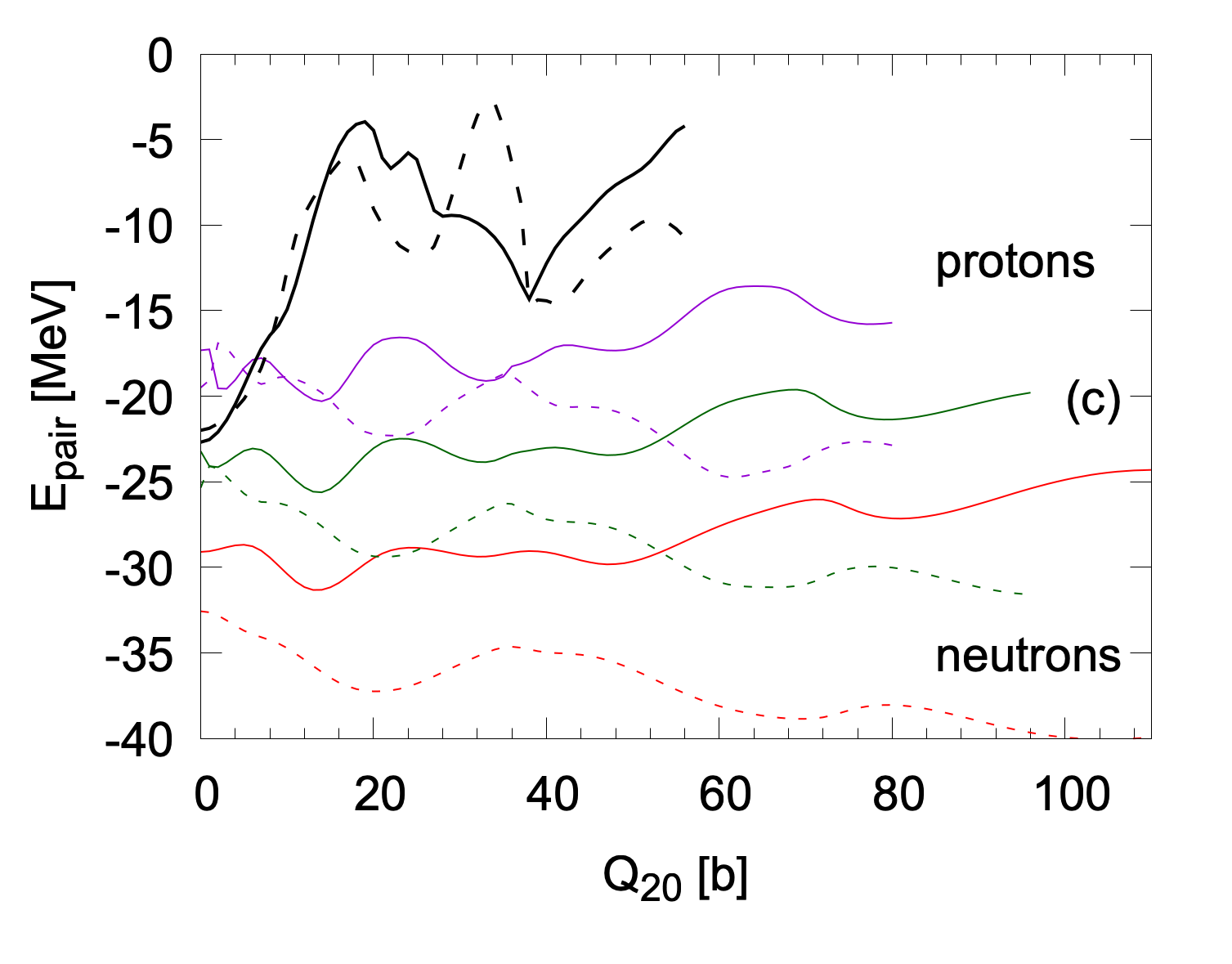}\\
\vskip-1.2cm
\hskip-0.5cm
\includegraphics[scale=0.138,angle=0]{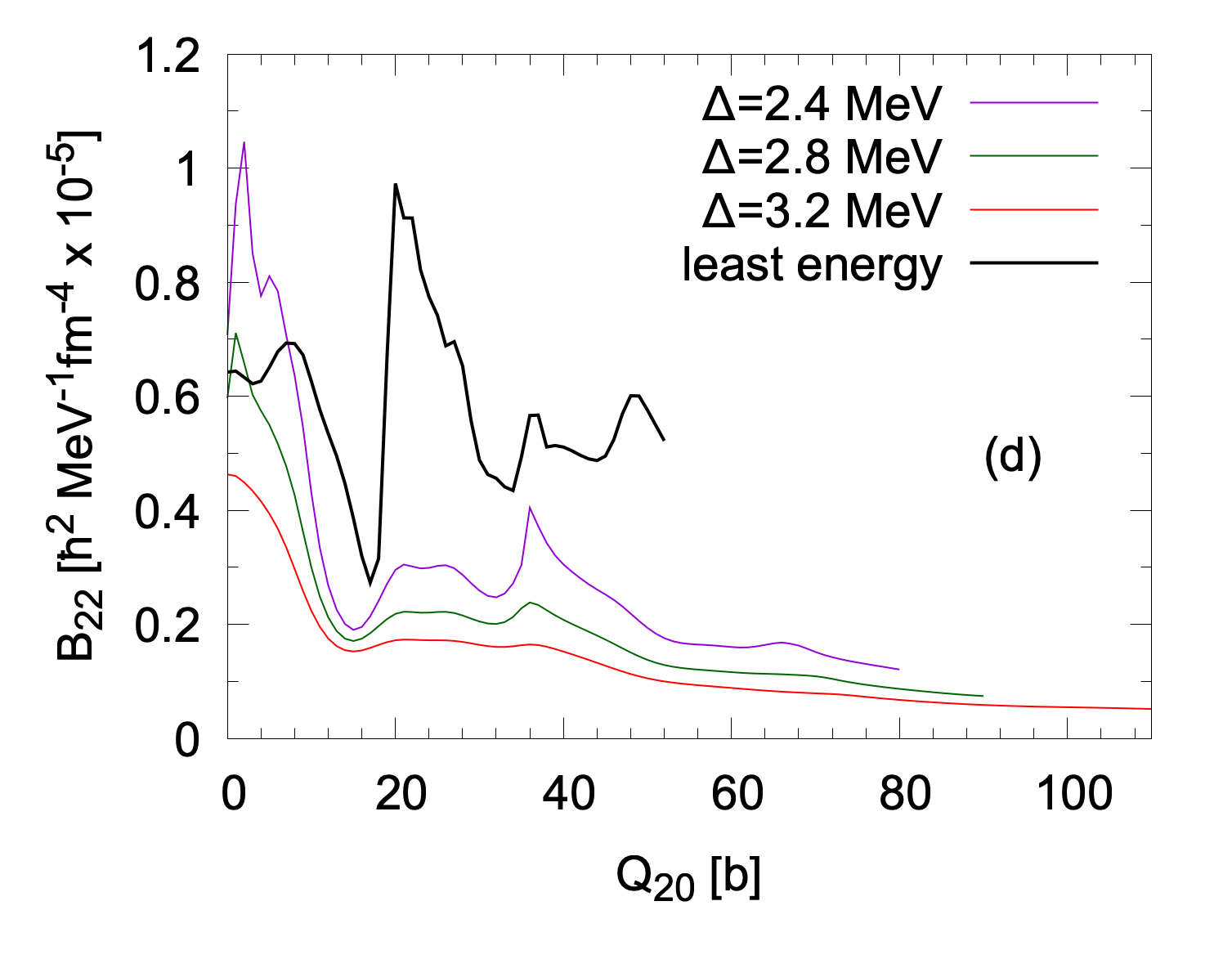}
\caption{(a) The PES (in MeV) of $^{262}$Rf isotope  in the ($Q_{20},\Delta$)
plane. The least-energy path is plotted with a white solid line. The
potential energy (b), the pairing energy (c), and the collective inertias (d)
along the least-energy path (black solid line) and for three constant values of the pairing gap.
}
\label{delt}
\end{figure}

\begin{figure}[h!]
\hskip-0.5cm
\includegraphics[scale=0.136,angle=0]{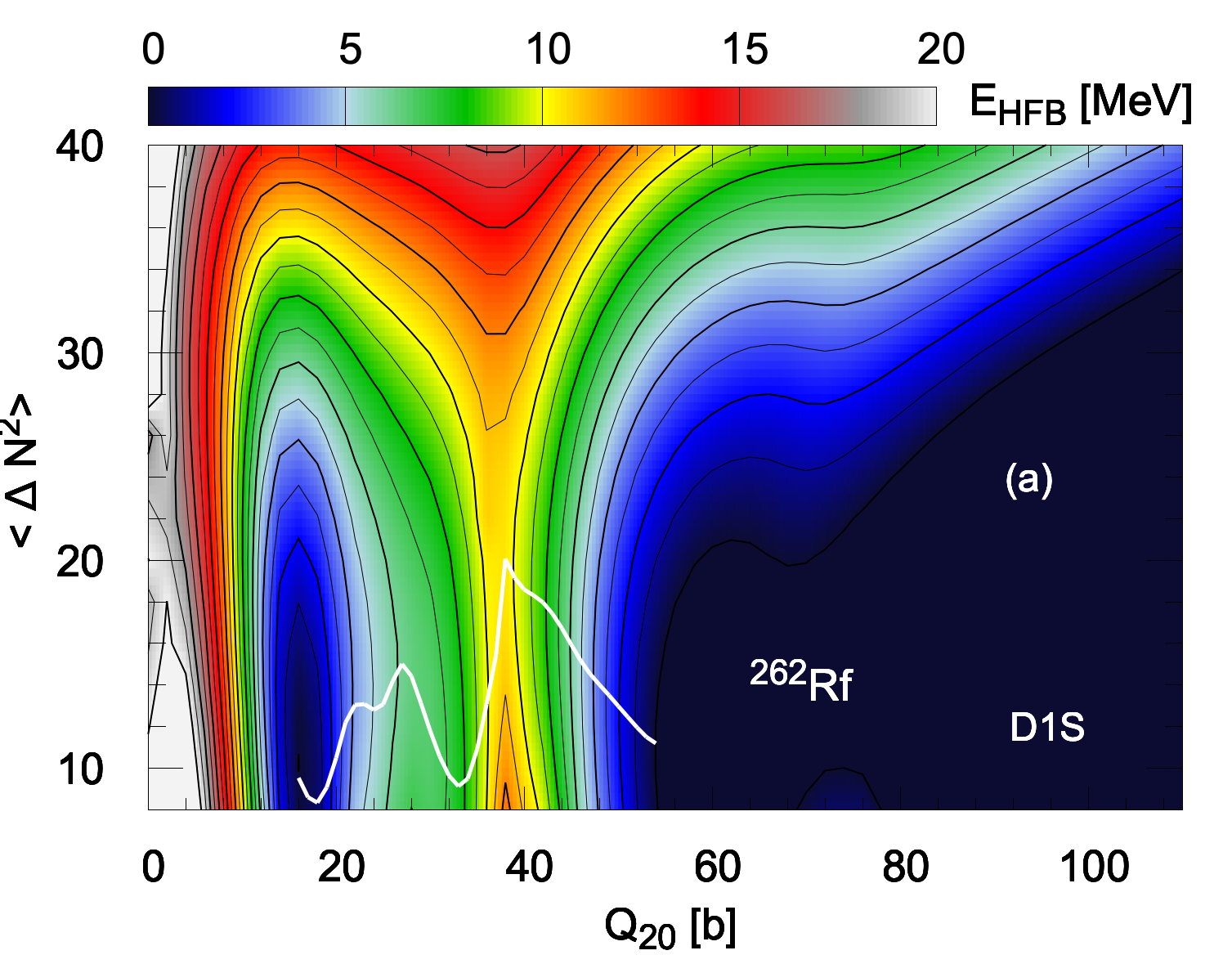}\\
\vskip-0.05cm
\hskip-0.5cm
\includegraphics[scale=0.138,angle=0]{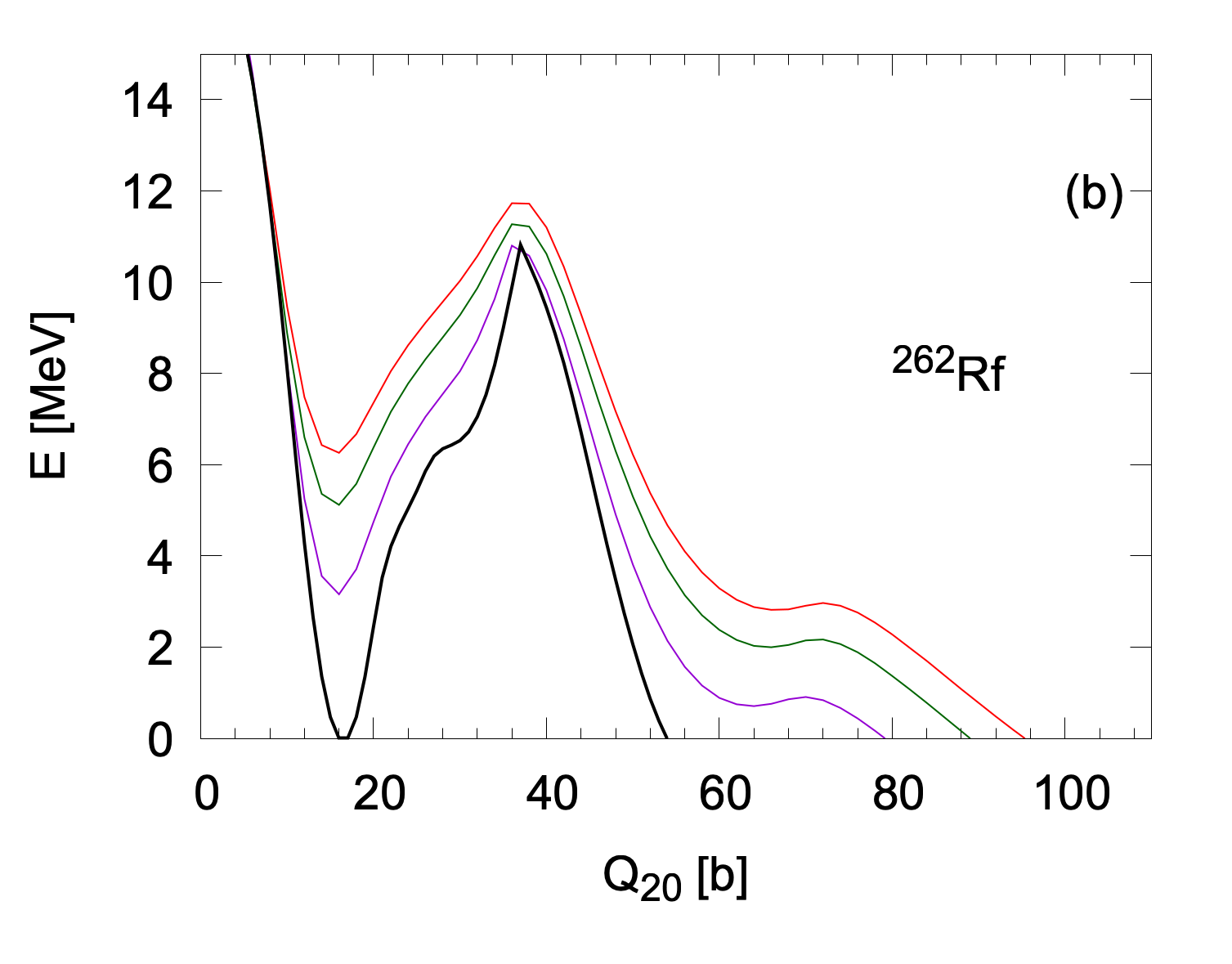}\\
\vskip-1.2cm
\hskip-0.5cm
\includegraphics[scale=0.138,angle=0]{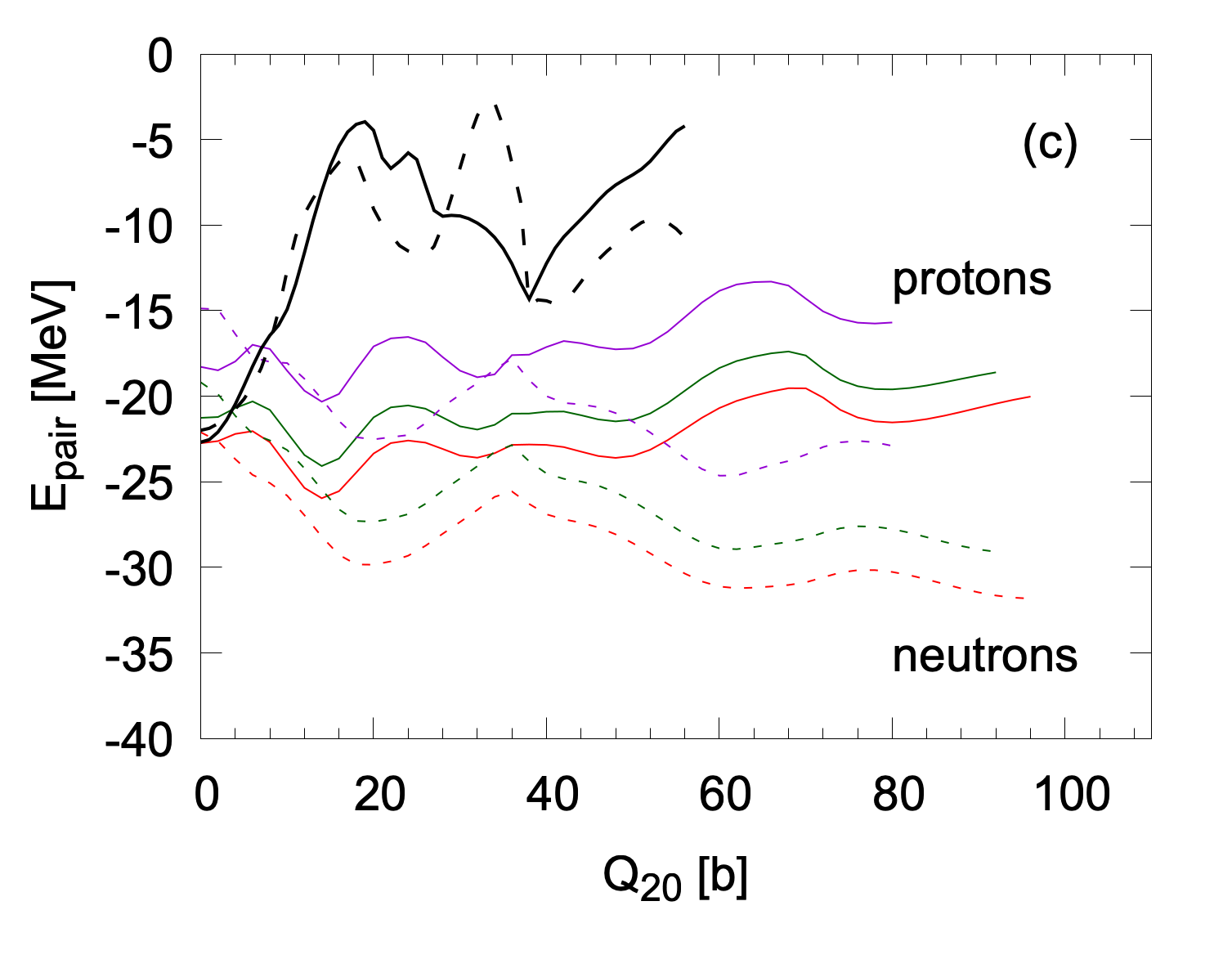}\\
\vskip-1.2cm
\hskip-0.5cm
\includegraphics[scale=0.138,angle=0]{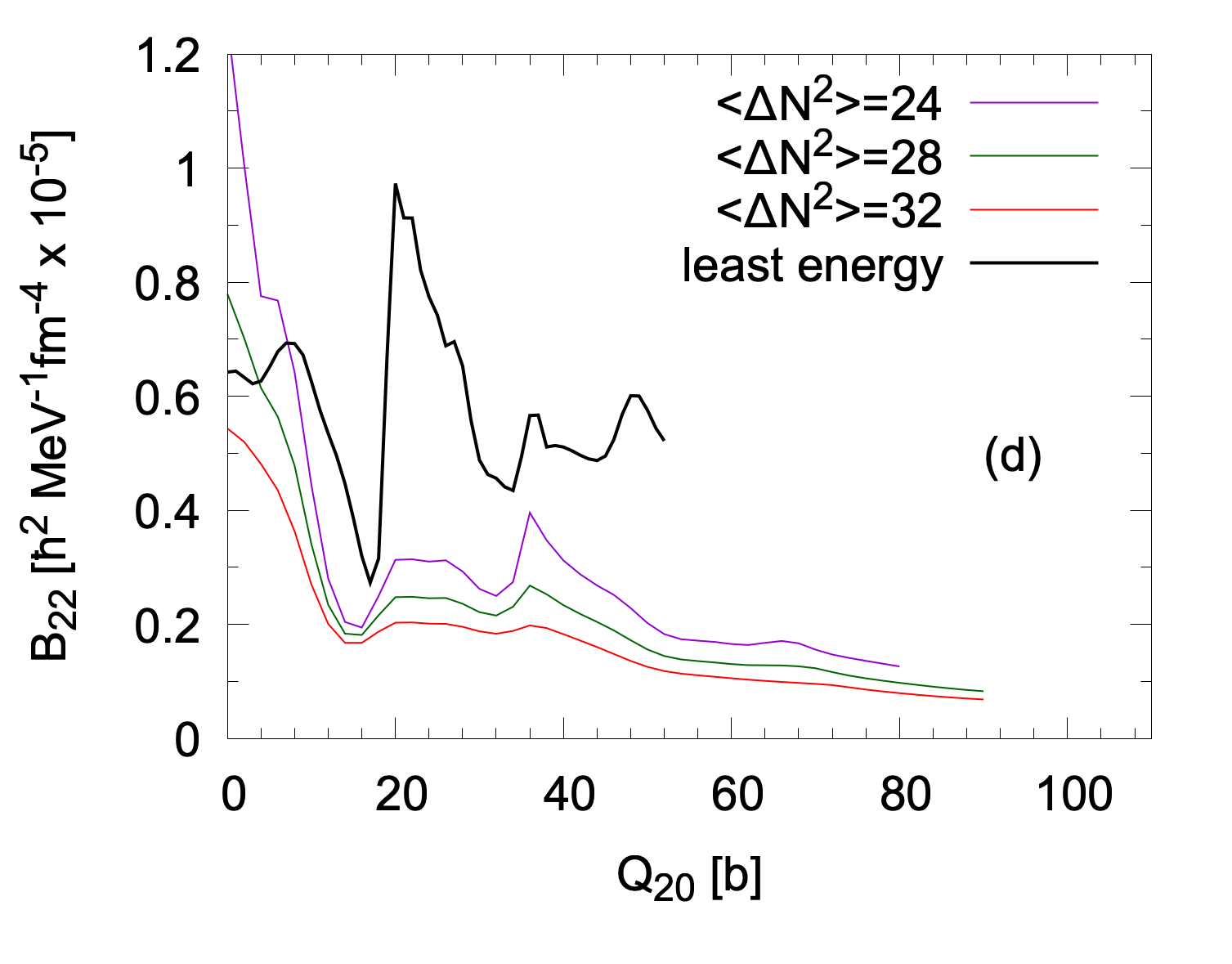}\\
\caption{Similar as Fig.~\ref{delt} but when $Q_{20}$ and $ \Delta
\hat{N}^2 $ are considered as collective degrees of freedom.}
\label{pnf}

\end{figure}

\begin{figure}[h!]
\includegraphics [scale=0.165] {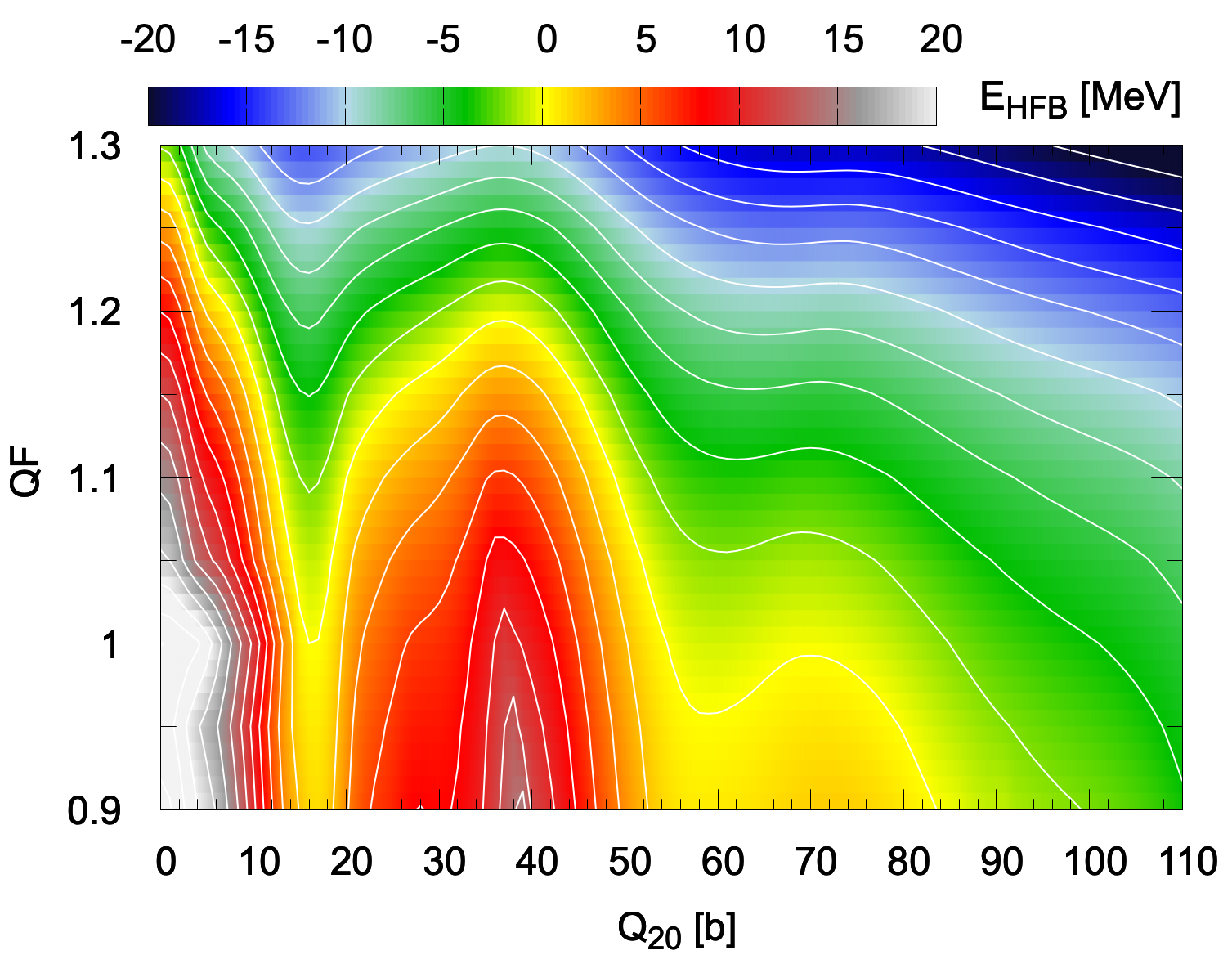}
\caption{The PES in quadrupole moment and the pairing strength factor 
along the least-energy path. The energy is plotted in relation to the 
ground state energy of the nucleus for QF=1.}
\label{qftot}
\end{figure}

\begin{figure*}[!htb]
     \hskip-0.4cm
    \includegraphics [scale=0.175] {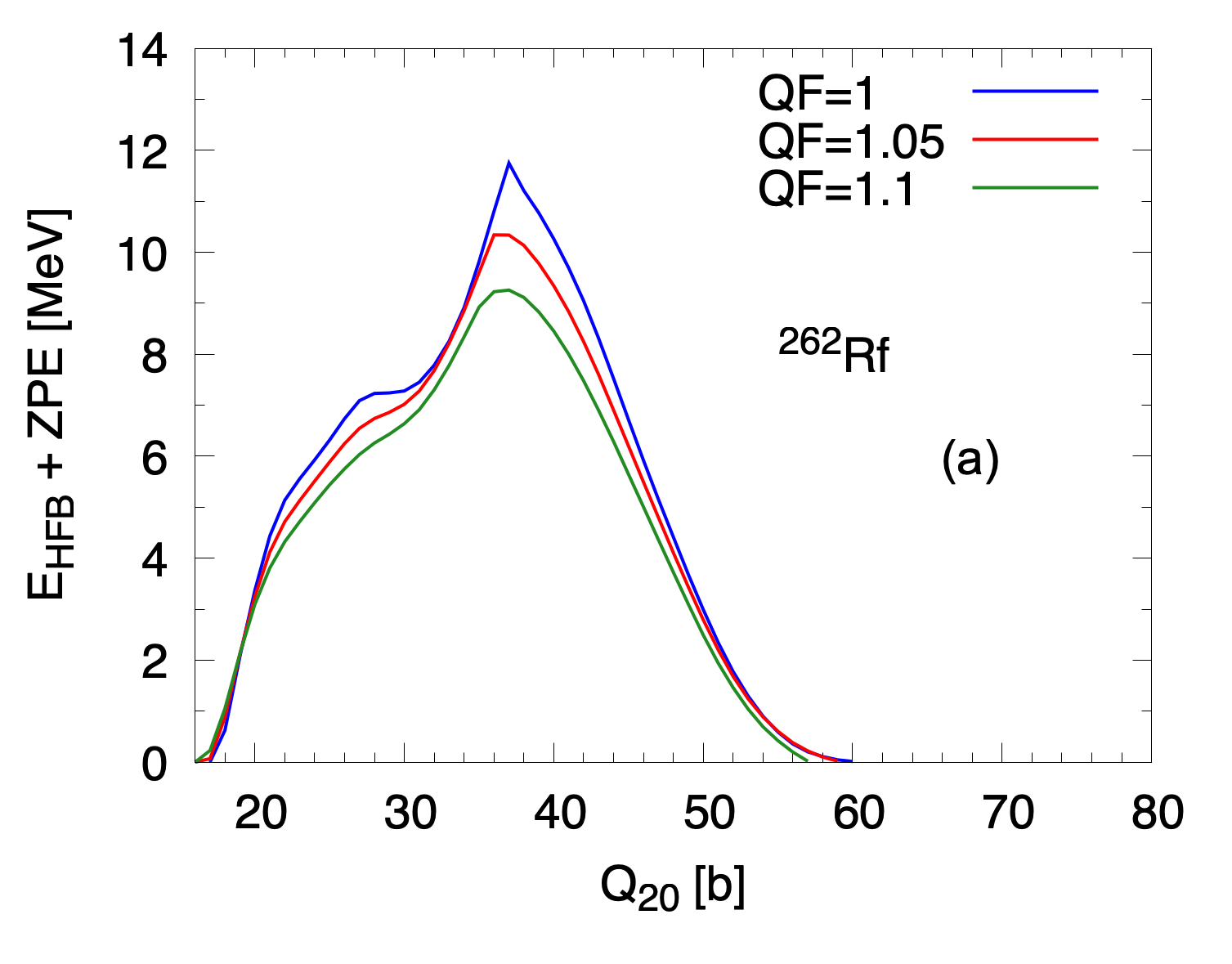}
     \hskip-0.4cm
        \includegraphics [scale=0.175] {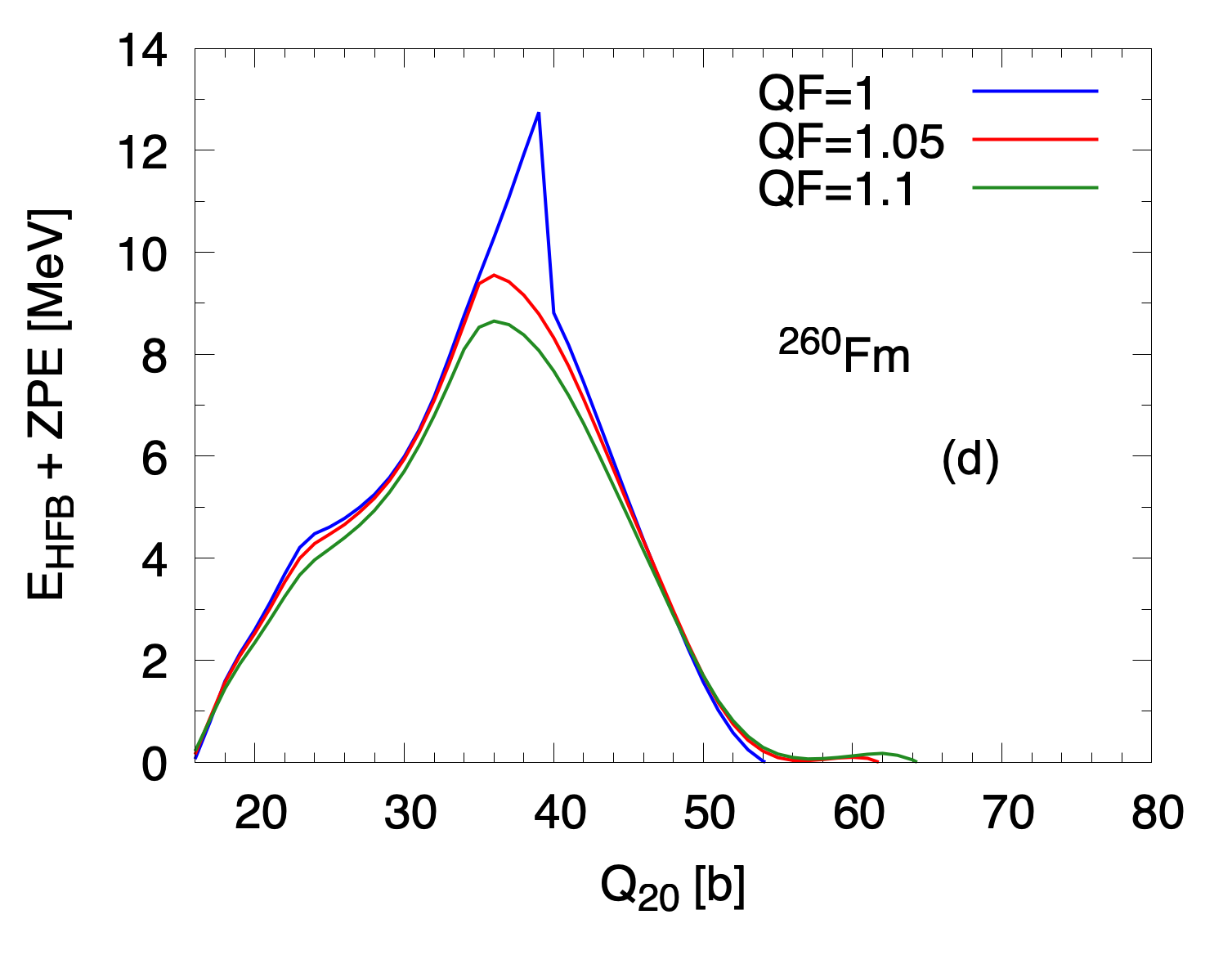}\\
             \vskip-0.cm
             \hskip-0.4cm
              \includegraphics [scale=0.175] {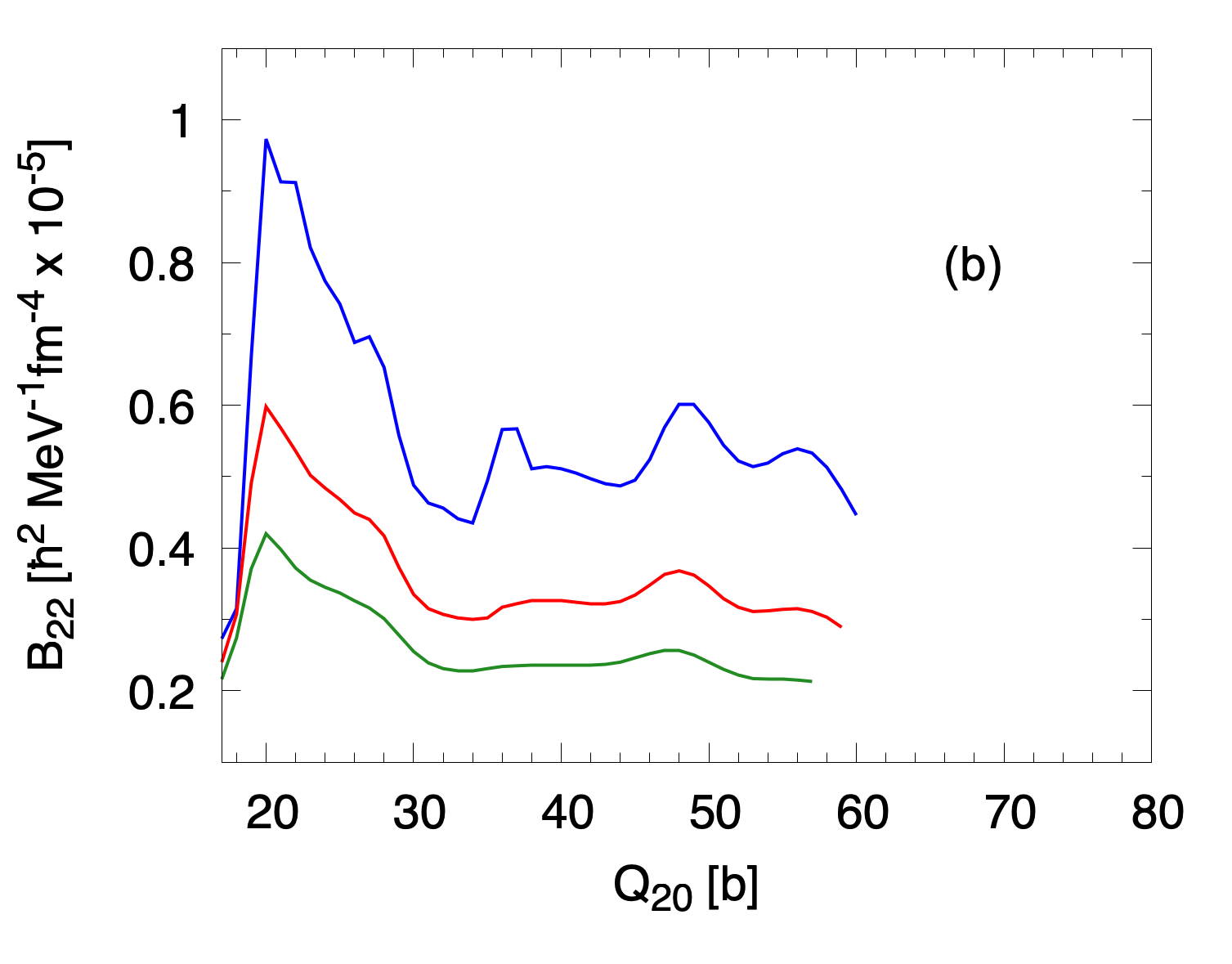}
                  \hskip-0.4cm
    \includegraphics [scale=0.175] {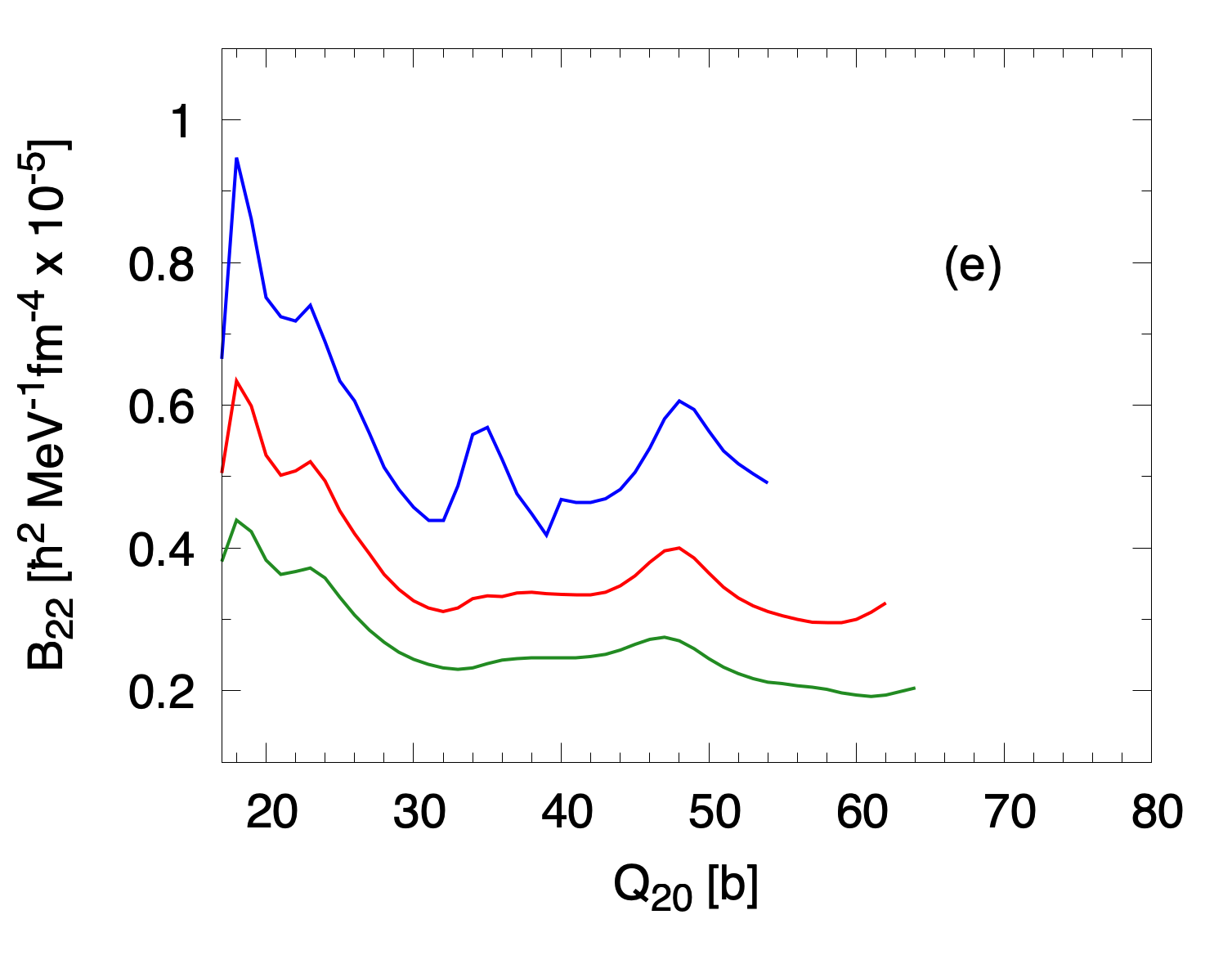}\\
         \vskip-0.cm
         \hskip-0.4cm
         \includegraphics [scale=0.175] {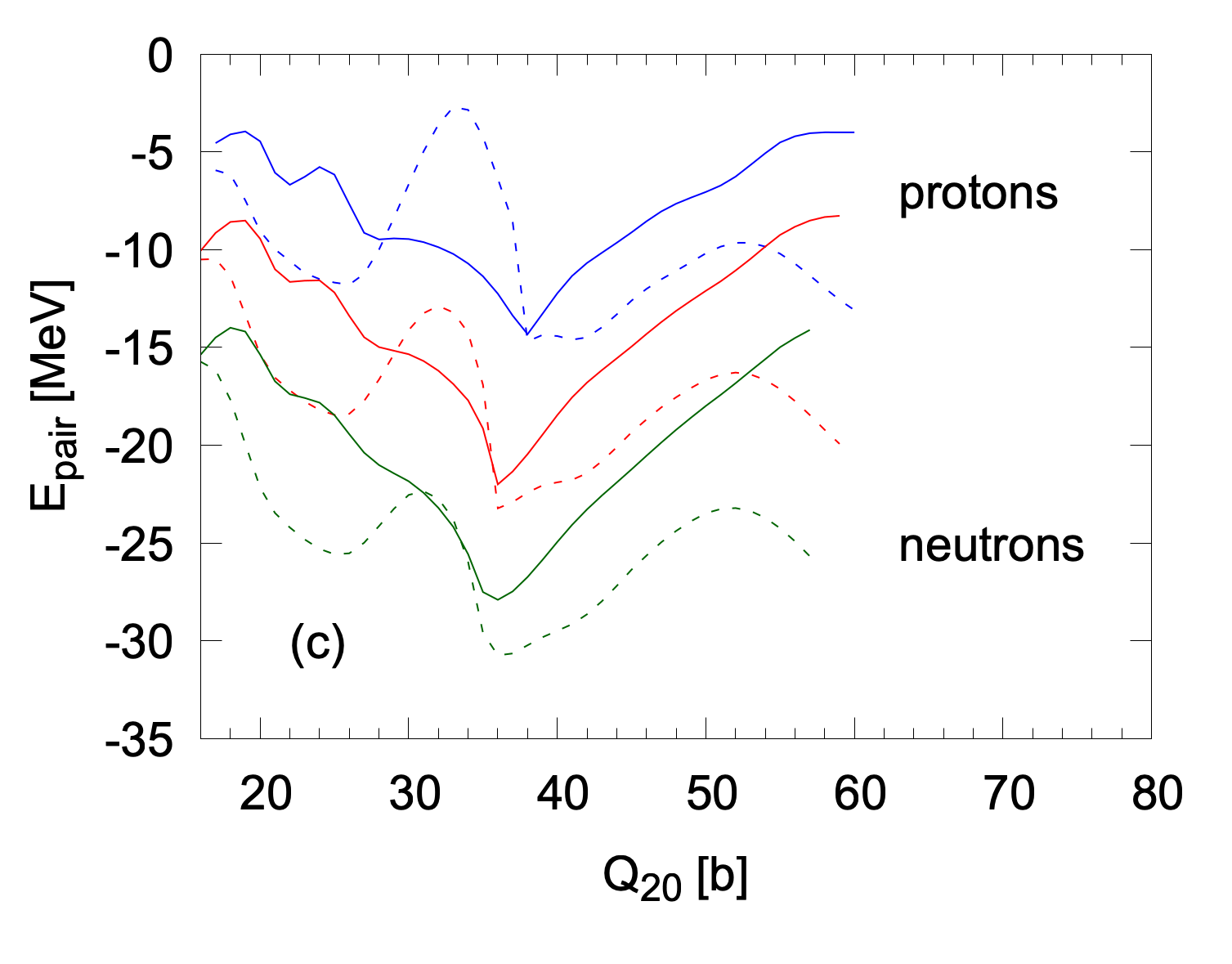}
                           \hskip-0.4cm
     \includegraphics [scale=0.175] {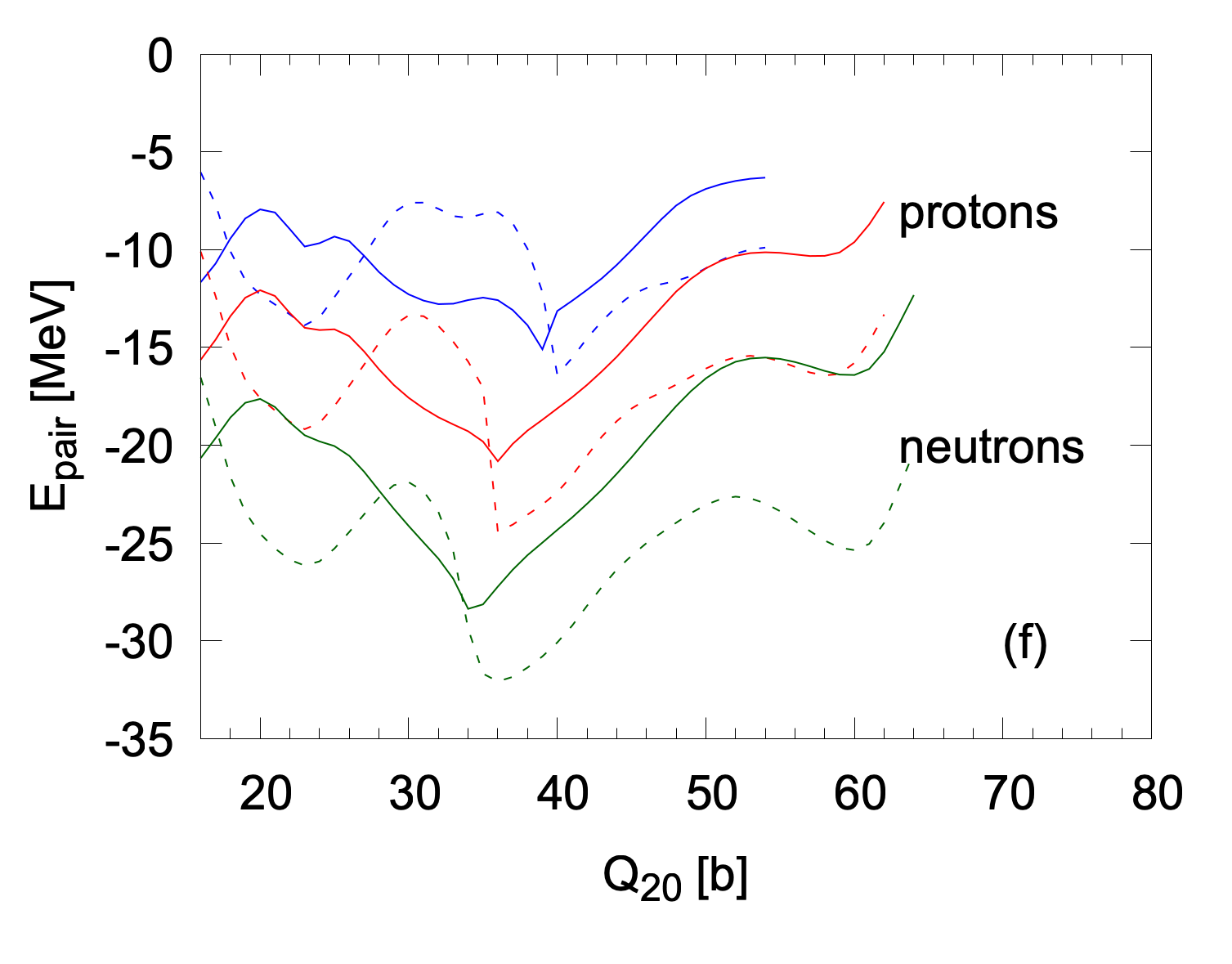}

	\caption{The HFB energy with the ZPE corrections (upper panel a,d), 
	the perturbative mass parameters (middle panel b,e) as a function 
	of $Q_{20}$ along the least-energy path and the corresponding 
	pairing energy (c), (f) of protons (solid line) and neutrons 
	(dashed line) for the $^{262}$Rf and $^{260}$Fm isotope.}
\label{merf}
\end{figure*}

\begin{figure}[h!]
\includegraphics [scale=0.22] {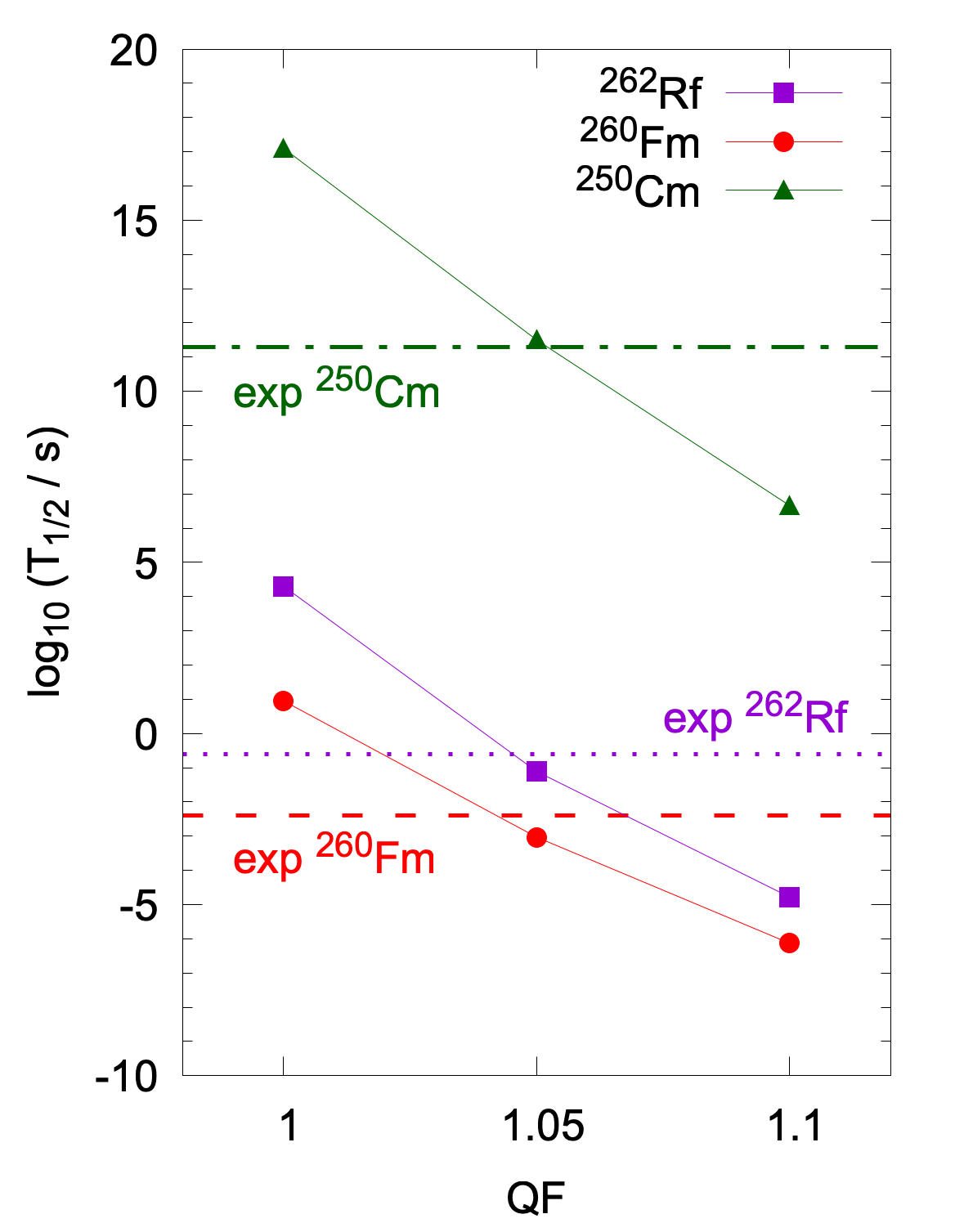}
\caption{The logarithm of spontaneous fission half-live of $^{250}$Cm,    
$^{260}$Fm and  $^{262}$Rf isotopes obtained within different values of 
the paring strength and compared with the experimental data taken 
from~\cite{Wapstra} depicted as horizontal lines.}
\label{tsfqf}
\end{figure}

\begin{figure}[h!]
\includegraphics [scale=0.17] {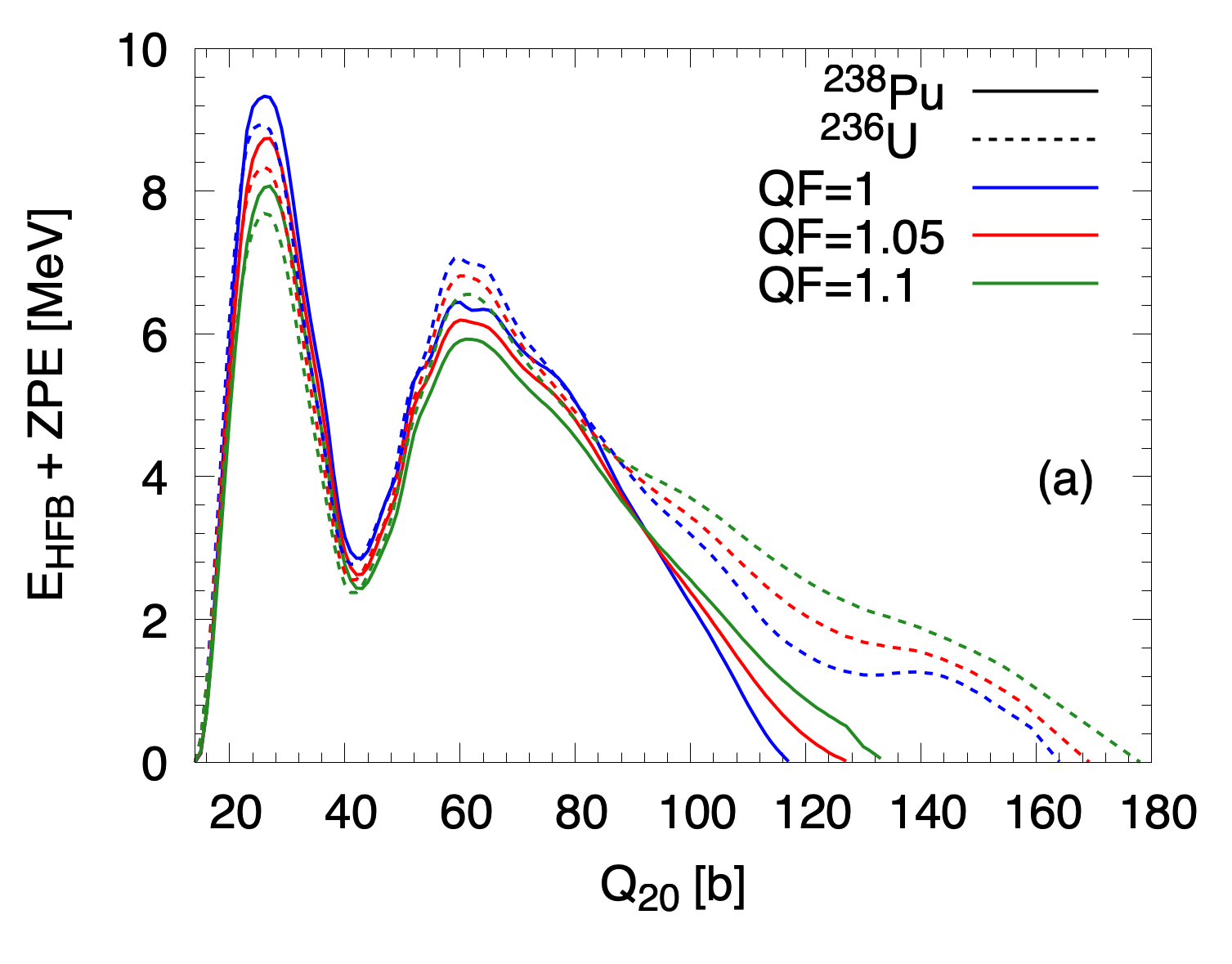}
\includegraphics [scale=0.17] {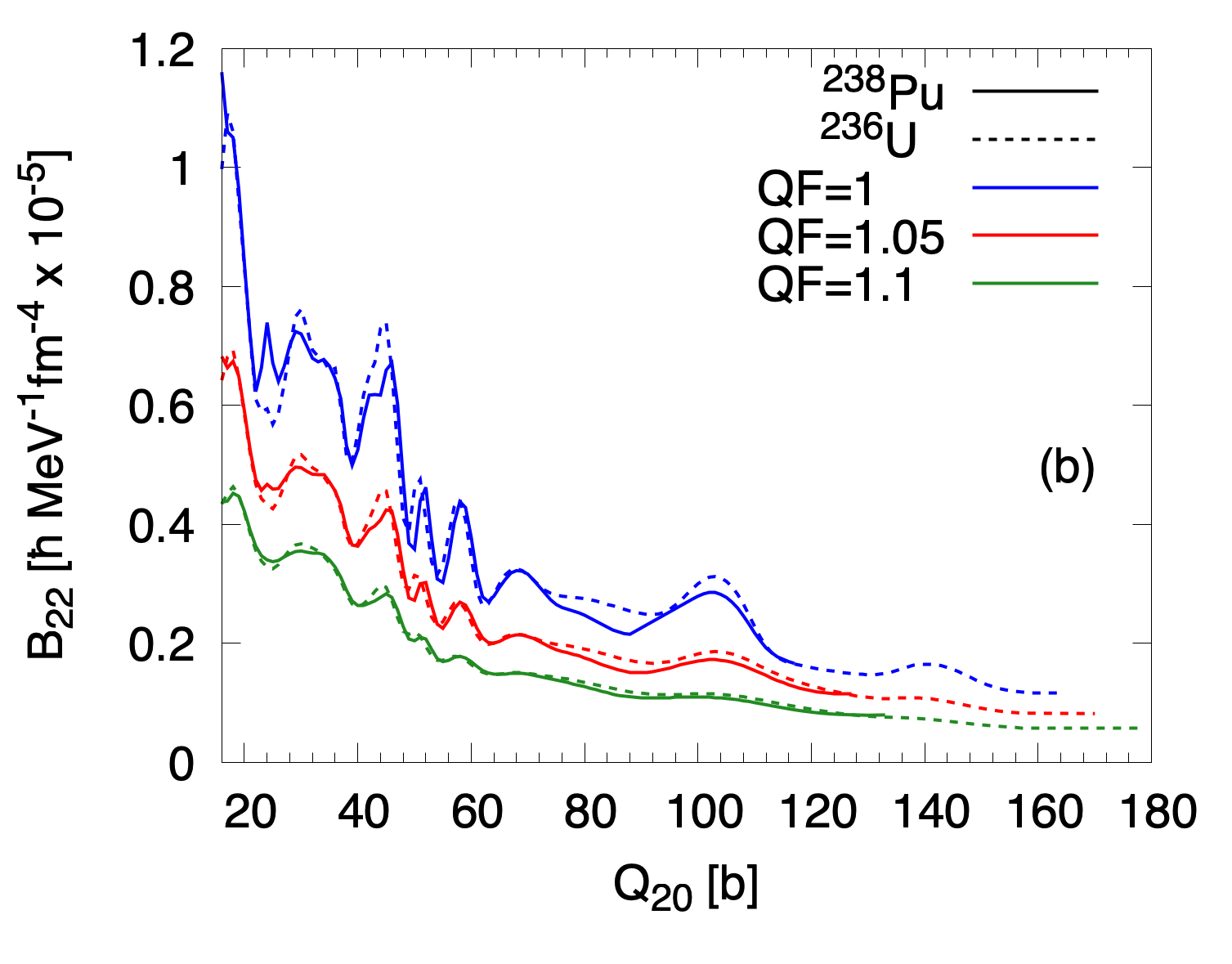}
\caption{(a) The least-energy path profiles of $^{238}$Pu (solid lines) 
and $^{236}$U (dashed lines) for different values of the quenching 
factor (QF) and (b) the corresponding collective inertias $B_{22}$ 
along the least-energy paths.}
\label{ePuU}
\end{figure}

\begin{figure*}[!htb]
     \includegraphics [scale=0.21] {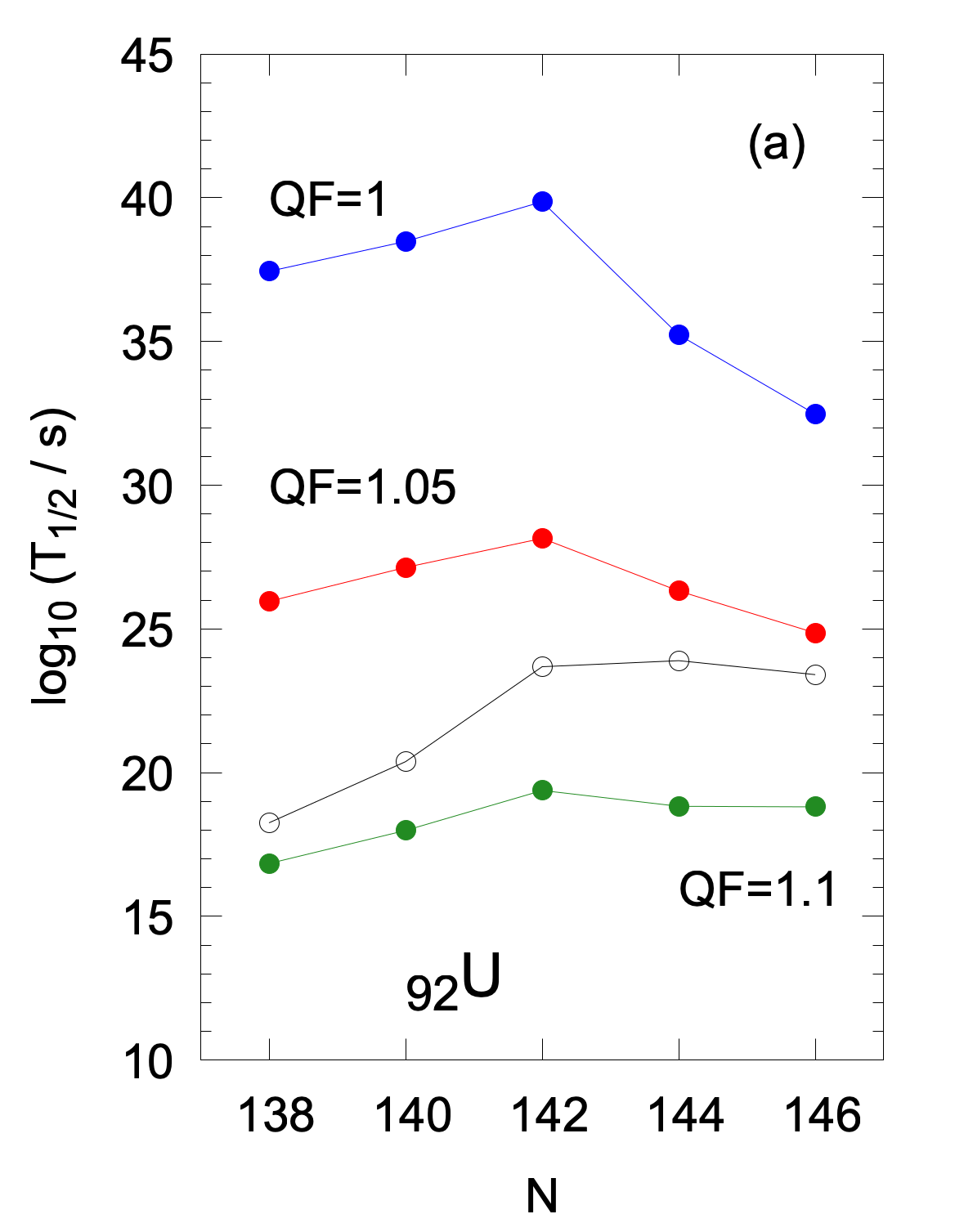}
          \includegraphics [scale=0.21] {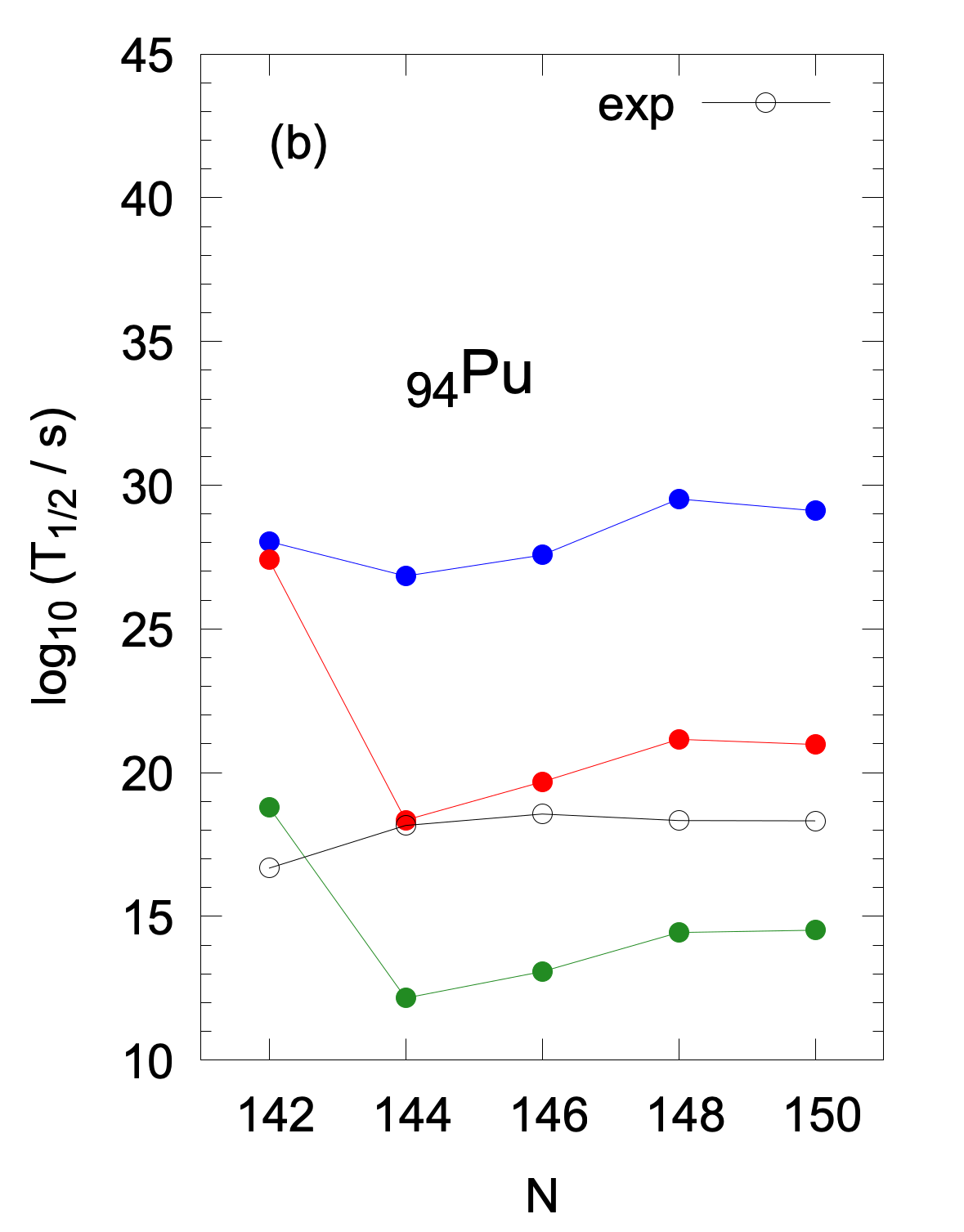}
    \caption{The logarithms of spontaneous fission half-lives of uranium (a)
    and plutonium (b) isotopes obtained within different values of the paring
    strength and compared with the experimental data (black circles).
}
\label{puu}\end{figure*}

Our first goal is to understand the impact of the pairing gap, 
particle number fluctuations, and the imposed pairing strength on the 
spontaneous fission static description. In order to clarify the picture 
and to minimize the influence of the secondary effects by involving parity breaking, we decided to 
focus first on the case of symmetric fission. We have chosen for our 
analysis the $^{262}$Rf isotope for which spontaneous fission 
half-life, as well as the fission fragment mass yields, have been measured 
and are available for comparison.

%%%%%%%%%%%%%%%%%%%%%%%%%%%%%%%%%%%%%%%%%%%%%%%%%
\subsection{The pairing gap}
%%%%%%%%%%%%%%%%%%%%%%%%%%%%%%%%%%%%%%%%%%%%%%%%%s

As has already been emphasized in the Introduction, the first studies 
within a simple, realistic model studying the influence of pairing 
in the description of fission showed that an increase in the pairing 
gap parameter leads to an increase in the penetrability of the fission 
barrier. This effect is strictly related to the fact that the 
collective inertias decrease as the inverse square of the pairing 
gap~\cite{babinet}. 

Panel (a) of Fig.~\ref{delt} depicts the potential 
energy surface (PES) in the ($Q_{20}, \Delta$) plane of $^{262}$Rf. The 
solid white line represents the least-energy path in this space of  
collective coordinates. To facilitate the understanding of the influence 
of $\Delta $ on the potential energy of the nucleus, we have also 
created one-dimensional plots of the potential energy (panel b) and 
pairing energy (panel c) as a function of nuclear elongation. The black 
line corresponds to the values along the least-energy path, while the other colors 
are associated with calculations with a constrained value of $\Delta$.
%represent changes in energy as a function of $Q_{20}$ for a fixed value 
%of the pairing gap greater than what was found in one-dimensional 
%calculations with the constraint only on the quadrupole moment. 
It is 
evident that as $ \Delta$ increases, the energy of the system increases 
as well. As expected, this is accompanied by an increase in the 
absolute value of pairing energy, as shown in panel (b), and a 
significant decrease in the values of collective inertias, as shown in 
panel (d). The decrease in the inertia follows quite closely the
inverse of the square of the pairing gap law as it can be checked that 
the inertias for $\Delta=2.4$ are a factor 1.8 larger than the ones of $\Delta=3.2$ 
over the whole range of quadrupole moments. Apart from that, one should also notice that the fission barrier 
becomes gradually broader for higher $\Delta$ values.

%

%%%%%%%%%%%%%%%%%%%%%%%%%%%%%%%%%%%%%%%%%%%%%%%%%
\subsection{The particles number fluctuation}
%%%%%%%%%%%%%%%%%%%%%%%%%%%%%%%%%%%%%%%%%%%%%%%%%

The particle number fluctuations $ \Delta \hat{N}^2 $, which is 
a quantity proportional to the pairing gap, have been successfully used 
to estimate the spontaneous fission half-lives of
actinides~\cite{PhysRevC.90.061304, Giuliani2014, PhysRevC.93.044315, PhysRevC.107.044307}.
Most of the studies found that the coupling of $ \Delta \hat{N}^2 $
with shape degrees of freedom greatly improves the agreement
between theoretical and experimental results. Now, we have decided to explore
in detail how the particle number 
fluctuations as a collective degree of freedom impact the 
fission barriers. The panel (a) of Fig.~\ref{pnf} 
presents the PES of $^{262}$Rf  in the ($Q_{20}, \Delta \hat{N}^2 
$) plane. One can notice that the topography of the presented surface looks very 
similar to Fig.~\ref{delt}. Also, the behavior of the 
analyzed quantities presented in panels (c) and (d) is similar to the 
one observed when the pairing gap was employed as a collective degree of freedom. This is a direct 
consequence of the almost linear connection between $\Delta$ and 
$ \Delta \hat{N}^2 $ variables (see Fig.~\ref{fit} in the Appendix). 
As $ \Delta \hat{N}^2 $ increases, the potential energy increases and
the collective inertia decreases, both following a quadratic trend as
can be observed in Fig 1 in Ref \cite{Giuliani2014}. 
It is also observed that an increase in the particle number fluctuations is 
accompanied by the broadening of the fission barrier.
As the energy and
collective inertia profiles 
obtained with $\Delta$ and $ \Delta \hat{N}^2 $ show a very 
similar pattern, it is possible to conclude that these two quantities may be treated 
interchangeably as collective variables. To test this conclusion, we
have performed calculations of the overlap between the HFB states obtained
with the constraint on $\Delta \hat{N}^{2}$ and that obtained with constraint
on $\Delta$ to the value obtained with the $\Delta \hat{N}^{2}$ constraint. The
values of the overlaps are always very close to one.

%%%%%%%%%%%%%%%%%%%%%%%%%%%%%%%%%%%%%%%%%%%%%%%%%
\subsection{The quenching factor\label{sec:qf_bf}}
%%%%%%%%%%%%%%%%%%%%%%%%%%%%%%%%%%%%%%%%%%%%%%%%%s

By considering different values for the quenching factor, one can scale 
the HFB proton and neutron pairing fields, which in practice leads to 
an increase or decrease of the  pairing strength with respect to the 
nominal value found in the process of fitting the parameters of the 
effective interaction.  It has been shown that by increasing the 
quenching factor by 10-20\%, one can obtain a better agreement with the 
data for the typical fission observables, like spontaneous fission 
half-lives of 
even-even~\cite{Giuliani2013,Rodriguez-Guzman2014,Rodriguez-Guzman2014a} 
and odd-mass \cite{guzman17} systems as well as fission fragment mass 
distributions \cite{PhysRevC.107.034307, PhysRevC.104.044612}. 
Fig.~\ref{qftot} depicts the least-energy path of $^{262}$Rf as a 
function of elongation for various values of the quenching factor. The 
presented PES is plotted in relation to the path obtained with 
$\textup{QF}=1$. Since the quenching factor is one of the three 
variables related to pairing, one would expect that changes in the 
value of this parameter would lead to similar effects on energy and 
collective inertias as in the previously analysed $ \Delta \hat{N}^2 $ 
and $\Delta$ cases. However, the results of the calculations conducted 
for the variable QF demonstrate that we are dealing with a different 
effect here, as we are changing the wave function by changing the 
underlying interaction. It should be noted that as the quenching factor 
increases, the potential energy of the nucleus decreases as the 
attractive pairing interaction becomes stronger. This is the opposite 
effect compared to what was observed earlier for increasing $\Delta 
\hat{N}^2$ and $\Delta$, which caused an increase in the potential 
energy. It is also worth noting that while increasing the particle 
number fluctuations, and the pairing gap results in higher and broader 
fission barrier, in the case of an increase in QF, the barrier becomes 
lower while its width does not change significantly. Furthermore, as 
shown in Fig.~\ref{qftot}, further increasing the value of QF results 
in a monotonic decrease in the potential energy of the nucleus and a 
flattening of the fission barrier. 

One can conclude that the quenching factor can be used as an artifact 
to renormalise locally the pairing strength in order to better 
reproduce experimental data. It can also be used for exploring fission 
properties with enhanced pairing along the fission path. However, this 
method does not allow for dynamical adjustment of pairing strength while  
searching for the least-action path.

%%%%%%%%%%%%%%%%%%%%%%%%%%%%%%%%%%%%%%%%%%%%%%%%%
\subsection{The quenching factor and spontaneous fission half-lives evaluation}
%%%%%%%%%%%%%%%%%%%%%%%%%%%%%%%%%%%%%%%%%%%%%%%%%s

The usual approach to describe the spontaneous fission process is to 
follow the least-energy path. This is a reasonable approximation that 
produces qualitative and often quantitative results in good agreement 
with experimental data. By changing the quenching factor, one can 
expect a slightly different topography of the PES, as discussed in 
Sec.~\ref{sec:qf_bf}. The upper panel of Fig.~\ref{merf} presents 
the fission barrier profiles corrected by the zero-point energy (ZPE) 
for $^{262}$Rf (left) and $^{260}$Fm (right). We notice a reduction of 
the fission barrier by a few MeV when the pairing strength is increased 
by 10\%. The collective inertias $B_{22}$ are plotted in the middle 
panels of Fig.~\ref{merf}. Their values decrease together with 
increasing the quenching factor and become smoother. With the stronger 
pairing, the difference in the $B_{22}$ values becomes less apparent. 
The bottom panel of Fig.~\ref{merf} presents the pairing energies of 
protons and neutrons. Obviously, the increase of the quenching factor 
leads to an increase in the absolute value of pairing energy. 
Fig.~\ref{tsfqf} presents the logarithms of spontaneous fission 
half-lives resulting from the static studies for various values of QF. 
As expected from the above analysis, the decrease of the fission 
barriers and collective inertias lead to an increase in the 
probability of tunneling the fission barrier, leading to shorter 
half-lives as the pairing strength grows. 

Let us now take a look at the actinides in order to have some insight into 
the pairing properties at the asymmetric fission path.  For our studies, we have 
chosen the plutonium and uranium isotopic chains, where
the asymmetric fission mode exists.
In the panel (a) of Fig.~\ref{ePuU}, the least-energy paths of
two representative isotopes, namely $^{238}$Pu and  $^{236}$U, are plotted for various
values of QF. Here, the first hump of the barrier is reflection symmetric, whereas 
the second one is octupolly deformed. 
One can see that the increased QF lowers the first symmetric fission barrier
by approximately 0.4 MeV per 0.05 change in the QF value.
The second minimum 
and the second asymmetric fission barrier show a slightly smaller influence 
of pairing strength and
the diminishing barrier height between the considered QF values 
does not exceed 0.2 MeV. 
This trend is inverted beyond the second barrier, at around 
$Q_{20}$ = 90 b and energy becomes higher for 
a higher value of the quenching factor.
The collective inertias related to the discussed fission paths are presented in the lower panel of Fig.~\ref{ePuU}.
Similarly, as for the previously discussed symmetric fission, $B_{22}$ is reduced and smoothened out for higher QF values.  It is worth 
mentioning that in these calculations, despite the second barrier in 
this region of actinides leads through the octupole shapes, the components of the 
inertia tensor associated with this degree of freedom have been neglected. 

The calculated spontaneous fission half-lives are 
plotted in Fig~\ref{puu}. 
Looking at the results, one can immediately notice that the optimal 
value of the pairing strength for reproducing experimental half-lives 
corresponds to a QF between 105\% and 110\% for the D1S parametrization. 
The effect of the increased tunneling probability when the quenching factor 
increases, and consequently the reduction of the spontaneous fission half-lives, 
mostly comes 
from the action integral of the first hump of the fission barrier. 
This is because of the strong reduction of collective inertias
with higher QF, accomplished by diminishing the potential energy. 
The influence of the asymmetric part of the barrier on the half-lives with increasing 
QF is less important. First, despite this part of the barrier being relatively broad, 
the collective inertia is much smaller than at the first barrier. Second, the 
reduction of the potential energy around the peak is compensated by an increase of the energy in the 
tail of the barrier.
It means that the action integral is insensitive to the energy changes introduced by 
the QF in the asymmetric part of the least-energy path. This suggests that QF has a 
stronger impact when parity is not broken. Thus, the overall trend of the shorter 
half-lives for higher QF in the actinides mainly comes from the response of the 
symmetric barrier on the pairing strength value.

% -----------------------------------------------------------------------
\subsection{Collective inertias with constraints on $\Delta$ and $\Delta \hat{N}^2$}
% -----------------------------------------------------------------------

\begin{figure}[h!]
\includegraphics[scale=0.17,angle=0]{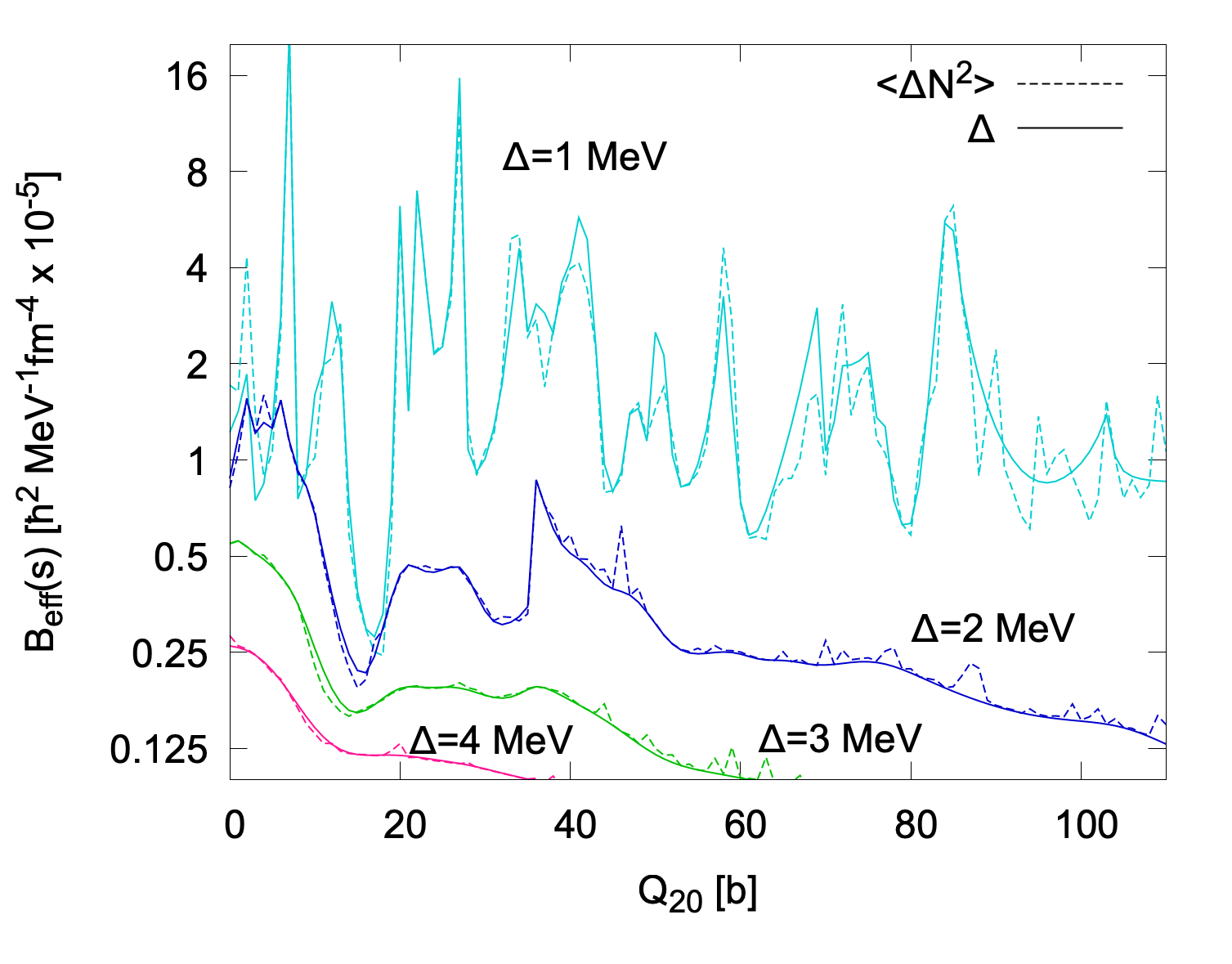}
\caption{ The effective inertia $B(s)$ along the quadrupole moment 
$s=Q_{20}$ for several values of $\Delta$ (solid line) and the 
corresponding $\Delta \hat{N}^2$ (dashed line) for $^{262}$Rf - see the 
text for the detailed explanation.
}
\label{inert}
\end{figure}

Further investigation of the role of microscopic pairing in fission 
should involve a dynamic approach requiring the calculation of the
collective inertia for the pairing modes.  Among the available quantities 
associated with the pairing correlations discussed in this paper, the 
quenching factor can not be considered a dynamical degree of freedom as 
it involves the change of the interaction itself, and it is, therefore
hardly consistent with the traditional frameworks to define collective
inertias. On the other hand, the  $\Delta$ and $\Delta \hat{N}^2$
collective variables can be used for that purpose, and it is
interesting to compare the inertias associated with both of them. As 
already mentioned in Section~\ref{THEORY}, 
$\Delta \hat{N}^2$ is a two-body operator, which means that 
the standard {\it cranking} approximation, defined for one body operators,  cannot be applied 
straightforwardly. Following the traditional procedure, we decided to check the collective 
inertia associated with this degree of freedom when one uses in the
formulas for the inertia of the quasiparticle one body part of $\Delta \hat{N}^2$.
In order to check the validity of this assumption, 
we compared the resulting $\Delta \hat{N}^2$ inertias with those
associated to $\Delta$. 
The effective inertia is defined in terms 
of the collective inertia tensor $B_{ij}$ along the fission path $s=Q_{20}$: 
\begin{equation} 
B_{\rm eff}(s)=\sum_{i,j}B_{ij}\frac{dq_i}{ds}\frac{dq_j}{ds}, 
\end{equation} 
where  $B_{ij}$ is the inverse of the collective mass tensor calculated 
within the perturbative {\it cranking} 
approximation~\cite{PhysRevC.84.054321}, and $(q_i, q_j)$ stand for the 
collective variable pair $(Q_{20}, \Delta)$ and $(Q_{20}, \Delta 
\hat{N}^2)$. We have considered several fixed values of $\Delta$ and 
related to their values of the $\Delta \hat{N}^2$ for which the 
collective inertias along the quadrupole moment have been computed. The 
results for $^{262}$Rf are shown in Fig.~\ref{inert}. The solid lines 
represent the inertias associated with $\Delta$ while the dashed lines 
correspond to the ones obtained when $\Delta \hat{N}^2$ is treated as 
a collective variable. As one can see, the effective inertias associated 
with $\Delta \hat{N}^2$ agree very well with the $\Delta$-dependent 
ones. It is worth emphasizing that the behavior of both inertias 
becomes more consistent as the value of $\Delta$ and $\Delta \hat{N}^2$ 
increases. This is particularly important in dynamic calculations, as 
one should expect that the path of the least-action path will primarily 
pass through higher values of these constraints. These results suggest 
that, in a first approximation, one can apply the one-body expression for 
the inertias in the perturbative {\it cranking} approximation employing 
the quasiparticle one body part of $\Delta \hat{N}^2$ as a dynamical 
variable. Developments in this direction are currently in progress.

%%%%%%%%%%%%%%%%%%%%%%%%%%%%%%%%%%%%%%%%%%%%%%%%%
\section{CONCLUSIONS\label{CONCLUSIONS}}
%%%%%%%%%%%%%%%%%%%%%%%%%%%%%%%%%%%%%%%%%%%%%%%%%

In this work, we analyzed the different methods traditionally employed in
studies exploring the impact of pairing correlations in the static description
of spontaneous fission. We compared the potential energy surfaces,
fission barrier, collective inertias, and pairing energies obtained when $\Delta$
and $\Delta \hat{N}^2$ are used as collective degrees of freedom.  The results
have been confronted with those obtained when a quenching factor QF is applied
to the HFB pairing field.

The obtained results of the microscopic studies of the impact of 
pairing on fission properties allow us to draw the following 
conclusions.

The amount of pairing correlations present along the fission path from ground state to 
scission can be controlled in various ways. One may increase the absolute value 
of the pairing energy by raising the pairing gap ($\Delta$),
the particle number fluctuations ($ \Delta \hat{N}^2 $), or
the pairing strength.

The increase of $\Delta$ and $ \Delta \hat{N}^2 $ leads 
to an increase of the potential energy, accompanied by a decrease of the 
collective inertias. In consequence, higher and broader fission barriers are obtained. 
Despite this, the smaller inertia reduces 
half-lives by a few orders of magnitude.

Conversely, the increase of the pairing strength by scaling the
quenching factor leads to a decrease of the potential energy. 
Lowering the fission barrier and the collective inertias simultaneously 
leads to a monotonic reduction of the fission half-lives for stronger pairing interactions. 
The effect of lowering the barrier with an increase of quenching 
factor is less significant in the region of the asymmetric fission 
channel. Thus, enhancing the pairing strength has a different 
impact on fission properties compared to the two previous techniques. This is because the nuclear 
interactions are modified. Changes in $\Delta$ and $ \Delta \hat{N}^2 $ 
affects only the single-particle level properties around the Fermi level, 
but the nuclear force remains unchanged.

The effective inertias associated with the $ \Delta \hat{N}^2 
$ computed within the ``one-body assumption" are in  very good 
agreement with the ones obtained for the pairing gap, which is a genuine
one body operator. This result gives a solid ground to the use
of $\Delta \hat{N}^{2}$ as collective variable in fission studies.

Static calculations of fission half-lives, taking into account various 
values of the quenching factor, do not exhaust the problem of 
describing spontaneous fission in terms of pairing forces. A more 
suitable treatment involves considering fission dynamics and should 
lead to a further reduction of computed spontaneous fission half-live.
Work along these lines is already in progress. A similar approach should be 
applied to study fission-fragments mass yields. This would require, 
however, the inclusion of the octupole moment as an additional degree of freedom, which 
increases the numerical complexity.

\begin{acknowledgements}
This research is part of the project No. 2021/43/P/ST2/03036 co-funded 
by the National Science Centre and the European Union’s Horizon 2020 
research and innovation programme under the Marie Skłodowska-Curie grant 
agreement no. 945339. The work of LMR and SAG
is supported by Spanish Agencia Estatal de Investigacion (AEI) of the 
Ministry of Science and Innovation (MCIN) under Grant No. 
PID2021-127890NB-I00. SG acknowledges support funded by
MCIN/AEI/10.13039/501100011033 and the ``European Union NextGenerationEU/PRTR''
under grant agreement No.~RYC2021-031880-I. 
\end{acknowledgements}

\appendix

\section{On the sign of occupation factor\label{APPENDIX}}

\begin{figure}[h!]
\includegraphics[scale=0.17,angle=0]{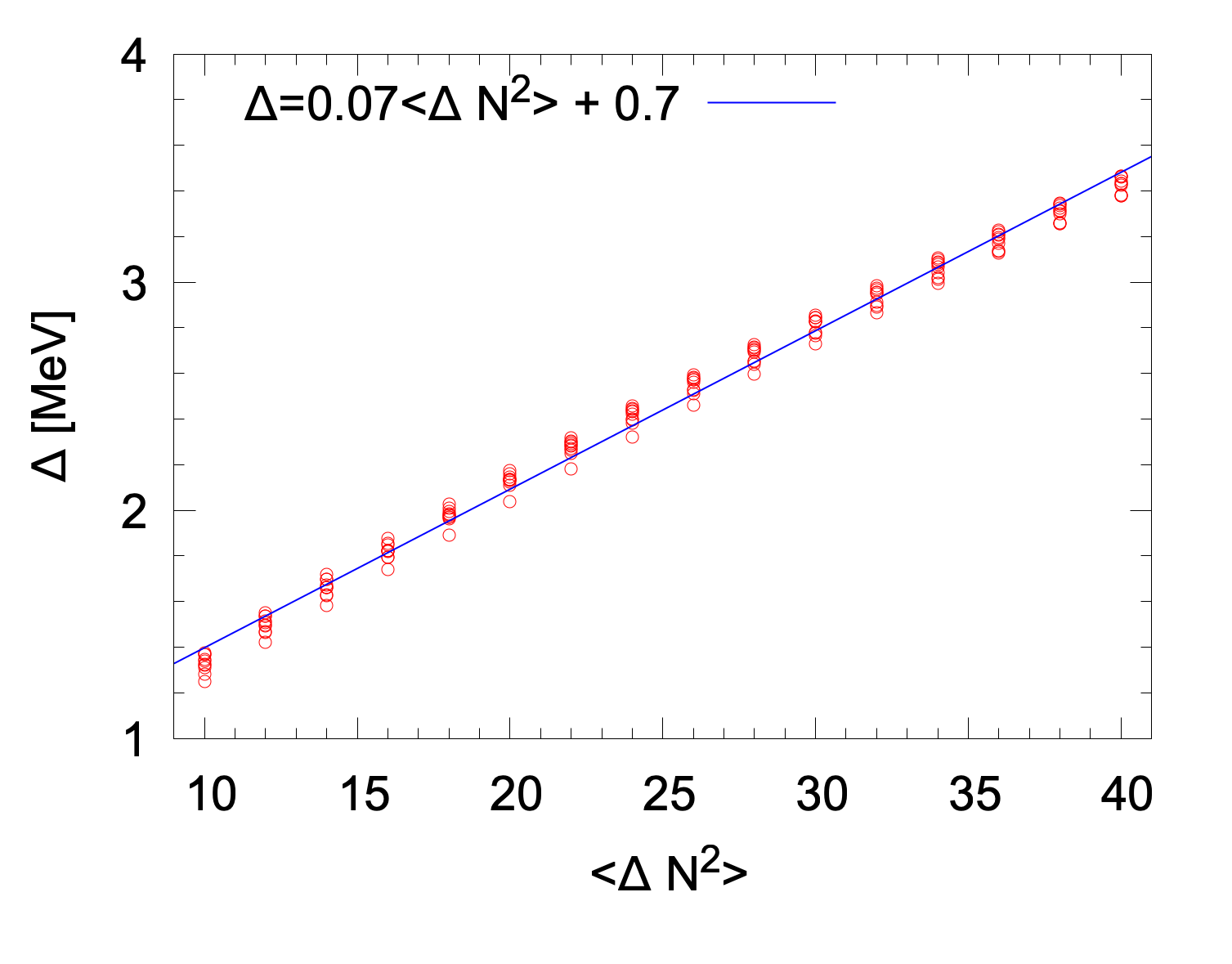}
\caption{ The relation between $\Delta$ and $\langle \Delta N^2\rangle$ for $^{262}$Rf.
The red circles correspond to the $\Delta$ values obtained in the
 calculations with the constrain on given $\langle\Delta N^2\rangle$ and
  the constrain on the various
$Q_{20}$ within the entire examined range (from 0 to 100 b).
}
\label{fit}
\end{figure}

The constraint in the gap parameter $\Delta=G\sum_k u_kv_k$ used in this 
paper presents a drawback 
associated with an overlooked property of the HFB and BCS theories. To 
discuss the problem, the BCS method will be used as the HFB one can 
be reduced to the former by working on the canonical basis of the density
matrix. In the standard BCS theory, the canonical transformation to 
quasiparticles is defined as:
\begin{align}
	\alpha_{k}^{+} &= u_{k}a^{+}_{k} - v_{k}a^{+}_{\bar{k}} \,, \\
	\alpha_{\bar{k}}^{+} &= u_{k}a^{+}_{\bar{k}} + v_{k}a^{+}_{k} \,,
\end{align}
where the variational parameters are assumed to be real and preserve 
time reversal invariance. The condition for BCS to be a canonical 
transformation requires that $u_{k}^{2}+v_{k}^{2}=1$. As the relative sign of the $u_{k}$ and 
$v_{k}$ coefficients in the BCS theory is not determined, traditionally, one chooses 
positive values for both sets of quantities. The most likely reason why 
the relative sign between $u$ and $v$ has never been considered is 
because pairing is often treated along with constant pairing 
interaction, for which all the relevant matrix elements are 
state-independent and attractive. In this specific case, the relative 
sign is irrelevant as the energy depends~\cite{RS80} upon powers of 
$v^2$ (density matrix) or in the parameter $\Delta$ defined before. As 
the $\Delta$ contribution to the energy is quadratic and attractive, 
the minimization principle favors the largest possible values of 
$\Delta$ and, therefore, coherence in the relative sign of $u$ and $v$ 
(as incoherent signs will average to zero in $\Delta$). The same argument 
also applies to state-dependent pairing, as the matrix elements are 
predominantly attractive. However, if pairing matrix elements depend on 
the quantum numbers of the involved orbitals (say $k$ or $l$), it might 
happen that some products $u_k v_k$ give a larger contribution to the 
energy than others,  $u_l v_l$. If we combine this possibility with a 
constrain in $\Delta$, it might happen that in order to increase the 
absolute value of $u_k v_k$ to gain energy and, at the same time, 
preserve the value of $\Delta$, some of the remaining terms $u_l v_l$ 
become negative. In our calculations preserving axial symmetry, $K$ 
(the projection of angular momentum along the $z$ axis) is a good 
quantum number, and all physical quantities, including the density and 
pairing tensor, are block diagonal in $K$. As a consequence, $\Delta$ 
receives independent contributions from each $K$ block, and we have 
witnessed in our calculations situations where all the $v_k$ of a given 
$K$ block change sign and give a negative contribution to $\Delta$. 
This configuration with the opposite sign of $v_k$ has lower energy 
than the one with all the $v_k$ having the same sign but obviously 
differ in the value of $\Delta$, and therefore they cannot be compared. 
In order to define $\Delta$ as a well-behaved collective variable, the 
sign of the $v_k$ has to be kept consistent for all the configurations. 
This difficulty does not show up obviously in the calculations where 
$ \Delta \hat{N}^2$ is constrained as the particle number 
fluctuation depends on the square of both $u_k$ and $v_k$ and the sign 
degree of freedom is not relevant to determine the value of the 
constraint. Therefore, in order to circumvent the problem with the constraint in $\Delta$
we have followed a simple receipt: Do the calculation with a constraint
on  $ \Delta \hat{N}^2$, change the relative sign between
$u_k$ and $v_k$ to be positive and compute the almost linear relation 
$\Delta (  \Delta \hat{N}^2 )$ to establish the correspondence
between $|\Phi ( \Delta \hat{N}^2) \rangle $ and $|\Phi (\Delta)\rangle $. 
Such a linear dependence of 
$\Delta (  \Delta \hat{N}^2 )$ for $^{262}$Rf in the relevant range
of $\Delta$ is presented in Fig.~\ref{fit}.

Please remember that the introduction of the constraint on $\Delta$ is to
have a constraint on a one-body operator where the traditional collective inertia 
formulas are justified, as opposed to the case of the two-body constraint
in $ \Delta \hat{N}^2$.

We point out that BCS wave function differing
in the sign of the $v_k$ no longer have overlap equals one. The overlap between
such wave functions is given by the product extended to all quantum numbers
in the basis of the factors $ u_{k}u'_{k}+v_{k}v'_{k}$. For those quantum
numbers where the occupancy $v'_{k}$ has the same sign as $v_{k}$, the factor
is $ u_{k}u_{k}+v_{k}v_{k} = 1 $. However, for those with opposite sign,
the factor becomes $ u_{k}u_{k} - v_{k}v_{k} \le 1 $, and therefore they
contribute to decreasing the overlap. 

%\section{REFERENCES}

%%%%%%%%%%%%%%%%%%%%%%%%%%%%%%%%%%%%%%%%%%%%%%%%%W
%\bibliographystyle{apsrev4-1}
%\bibliography{references}
\input{pair_stat_final.bbl}
%%%%%%%%%%%%%%%%%%%%%%%%%%%%%%%%%%%%%%%%%%%%%%%%%

\end{document}